\documentclass[a4paper,11pt]{article}
\pdfoutput=1

\usepackage[svgnames]{xcolor} 

\usepackage[printonlyused,withpage]{acronym} 
\usepackage{amsfonts}
\usepackage{anyfontsize}
\usepackage{bbm}
\usepackage{booktabs} 
\usepackage{braket}
\usepackage{cancel}
\usepackage{changepage}
\usepackage{enumitem}
\usepackage[mathscr]{euscript}
\usepackage[T1]{fontenc} 
\usepackage{gensymb}
\usepackage{Packages/jheppub}
\usepackage{listings} 
\usepackage{lmodern} 
\usepackage{makecell}
\usepackage{mathtools}
\usepackage[framemethod=tikz]{mdframed} 
\usepackage{microtype} 
\usepackage{multirow}
\usepackage{physics}
\usepackage{relsize}
\usepackage{rotating} 
\usepackage{sansmath}
\usepackage{simplewick}
\usepackage{slashed}
\usepackage{stmaryrd}
\usepackage{subcaption}
\usepackage{tcolorbox}
\usepackage{theorem}
\usepackage{tikz}
\usepackage{tikz-feynman,contour}
\usepackage[normalem]{ulem}
\usepackage{xspace} 
\usepackage{yfonts}
\usepackage[vcentermath]{youngtab}

\tcbuselibrary{theorems}
\usetikzlibrary{decorations.markings}

\def\d{{\mathrm{d}}} 

\def\ii{{\text{i}}}

\def\doot{{\boldsymbol{\hspace{0.1em} \cdot\hspace{0.1em}}}}

\makeatletter
\newcommand*{\transpose}{%
  {\mathpalette\@transpose{}}%
}
\newcommand*{\@transpose}[2]{%
  \raisebox{\depth}{$\m@th#1\intercal$}%
}
\makeatother

\newtcbox{\sln}{colback=Gainsboro,
colframe=Gainsboro}

\newcommand{\bt}[1]{{\sansmath{\boldsymbol{#1}}}}

\newcommand{\overbar}[1]{\mkern 2mu\overline{\mkern-4mu#1\mkern-4mu}\mkern 2mu}

\tikzset{snake it/.style={decorate, decoration={snake,amplitude=10mm}}}

\tikzset{/pgf/decoration/.cd,
    number of sines/.initial=10,
    angle step/.initial=20,
}

\newdimen\tmpdimen

\pgfdeclaredecoration{full sines}{initial}
{
    \state{initial}[
        width=+0pt,
        next state=move,
        persistent precomputation={
            \pgfmathparse{\pgfkeysvalueof{/pgf/decoration/angle step}}%
            \let\anglestep=\pgfmathresult%
            \let\currentangle=\pgfmathresult%
            \pgfmathsetlengthmacro{\pointsperanglestep}%
                {(\pgfdecoratedremainingdistance/\pgfkeysvalueof{/pgf/decoration/number of sines})/360*\anglestep}%
        }] {}
    \state{move}[width=+\pointsperanglestep, next state=draw]{
        \pgfpathmoveto{\pgfpointorigin}
    }
    \state{draw}[width=+\pointsperanglestep, switch if less than=1.25*\pointsperanglestep to final, 
        persistent postcomputation={
        \pgfmathparse{mod(\currentangle+\anglestep, 360)}%
        \let\currentangle=\pgfmathresult%
    }]{%
        \pgfmathsin{+\currentangle}%
        \tmpdimen=\pgfdecorationsegmentamplitude%
        \tmpdimen=\pgfmathresult\tmpdimen%
        \divide\tmpdimen by2\relax%
        \pgfpathlineto{\pgfqpoint{0pt}{\tmpdimen}}%
    }
    \state{final}{
        \ifdim\pgfdecoratedremainingdistance>0pt\relax
            \pgfpathlineto{\pgfpointdecoratedpathlast}
        \fi
   }
}

\pgfdeclaredecoration{completely sines}{initial}
{
    \state{initial}[
        width=+0pt,
        next state=upsine,
        persistent precomputation={\pgfmathsetmacro\matchinglength{
            \pgfdecoratedinputsegmentlength / int(\pgfdecoratedinputsegmentlength/\pgfdecorationsegmentlength)}
            \setlength{\pgfdecorationsegmentlength}{\matchinglength pt}
        }] {}
    \state{upsine}[width=\pgfdecorationsegmentlength,next state=downsine]{
        \pgfpathsine{\pgfpoint{0.25\pgfdecorationsegmentlength}{0.5\pgfdecorationsegmentamplitude}}
        \pgfpathcosine{\pgfpoint{0.25\pgfdecorationsegmentlength}{-0.5\pgfdecorationsegmentamplitude}}
    }
    \state{downsine}[width=\pgfdecorationsegmentlength,next state=upsine]{
        \pgfpathsine{\pgfpoint{0.25\pgfdecorationsegmentlength}{-0.5\pgfdecorationsegmentamplitude}}
        \pgfpathcosine{\pgfpoint{0.25\pgfdecorationsegmentlength}{0.5\pgfdecorationsegmentamplitude}}
}
    \state{final}{}
}

\tikzfeynmanset{ with arrow/.style = {
   decoration={
     markings,
     mark=at position 0.5
          with {\arrow[xshift=1.9mm]{Latex[width=2.75mm,length=3mm]}}
     },
   postaction=decorate}
}

\tikzfeynmanset{ with glarrow/.style = {
   decoration={
     markings,
     mark=at position 0.5
          with {\arrow[white,xshift=3mm]{Latex[width=4.125mm,length=4.5mm]}}
     },
   postaction=decorate}
}

\tikzfeynmanset{ with outarrow/.style = {
   decoration={
     markings,
     mark=at position 0.5
          with {\arrow[xshift=3.35mm]{Latex[width=4.58333mm,length=5mm]}}
     },
   postaction=decorate}
}

\tikzfeynmanset{ bigphoton/.style={
    /tikz/draw=none,
    /tikz/decoration={name=none},
    /tikz/postaction={
      /tikz/draw,
      /tikz/decoration={
        completely sines,
        segment length=3mm,
        amplitude=2.5mm,
      },
      /tikz/decorate=true,
    },
  },
}

\tikzfeynmanset{ roundbigphoton/.style={
    /tikz/draw=none,
    /tikz/decoration={name=none},
    /tikz/postaction={
      /tikz/draw,
      /tikz/decoration={
        full sines,
        segment length=3mm,
        amplitude=1.25mm,
      },
      /tikz/decorate=true,
    },
  },
}

\tikzset{ mega thick/.style= {line width = 3.4pt}
}

\hypersetup{colorlinks, breaklinks, linkcolor=black,citecolor=black,filecolor=black,urlcolor=black} 

\makeatletter
\renewcommand{\fnum@figure}{\textsc{\figurename~\thefigure}} 
\makeatother

\lstnewenvironment{pseudoc}{\lstset{frame=lines,language=C,mathescape=true}}{}
\lstnewenvironment{logs}{\lstset{frame=lines,basicstyle=\footnotesize\ttfamily,numbers=none}}{}
\lstnewenvironment{cc}{\lstset{frame=lines,language=C}}{}
\lstnewenvironment{c64}{\lstset{backgroundcolor=\color{c64},basewidth=0.65em,basicstyle=\commodoreface\color{c64light},numbers=none,framerule=10pt,rulecolor=\color{c64light},frame=tb,framexbottommargin=30pt}}{}
\lstnewenvironment{html}{\lstset{frame=lines,language=html,numbers=none}}{}
\lstnewenvironment{pseudo}{\lstset{frame=lines,mathescape=true,morekeywords={learn_string_domain, save_model}}}{}
\lstnewenvironment{pseudoctiny}{\lstset{language=C,mathescape=true,basicstyle=\tiny\sffamily}}{}
\lstnewenvironment{cctiny}{\lstset{language=C,basicstyle=\tiny\sffamily}}{}
\lstnewenvironment{pseudotiny}{\lstset{mathescape=true,basicstyle=\tiny\sffamily}}{}

\title{Exploring color-octet scalar parameter space in minimal \emph{R}-symmetric models}

\author{Linda M. Carpenter,}
\author{Taylor Murphy,}
\author{and Matthew J. Smylie}
\affiliation{Department of Physics, The Ohio State University\\
191 W. Woodruff Ave,\\ Columbus, OH 43210}

\emailAdd{lmc@physics.osu.edu}
\emailAdd{murphy.1573@osu.edu}
\emailAdd{smylie.8@osu.edu}

\date{\today}

\abstract{\begin{abstract}

In this work we study the collider phenomenology of color-octet
scalars (\emph{sgluons}) in minimal supersymmetric models endowed with a global continuous $R$ symmetry. We systematically catalog the significant decay channels of scalar and pseudoscalar sgluons and identify novel features that are natural in these models. These include decays in nonstandard diboson channels, such as to a gluon and a photon; three-body decays with considerable branching fractions; and long-lived particles with displaced vertex signatures. We also discuss the single and pair production of these particles and show that they can evade existing constraints from the Large Hadron Collider, to varying extents, in large regions of reasonable parameter space. We find, for instance, that a $725\, \text{GeV}$ scalar and a $350\, \text{GeV}$ or lighter pseudoscalar can still be accommodated in realistic scenarios.
\end{abstract}}

\begin{document}

\maketitle
\flushbottom

\section{Introduction}
\label{s1}

Supersymmetry (SUSY) remains one of the leading extensions of the Standard Model, as it provides a framework for stabilizing the weak scale, identifying a dark matter candidate, achieving gauge coupling unification, and solving other problems of theoretical and experimental interest. Nevertheless, the continuing absence of SUSY phenomena at colliders renders a look beyond the Minimal Supersymmetric Standard Model (MSSM) increasingly necessary. There exist numerous interesting enhancements of the MSSM, among the most promising of which are models with a global continuous $R$ symmetry \cite{Hall:1991r1, Randall:1992r2, Kribs:2008rs}.

In $R$-symmetric models, Majorana mass terms for the gauginos are forbidden; instead, gauginos obtain their masses through Dirac couplings to new fermions. Minimal models must therefore feature new chiral superfields that transform in the adjoint representation of each Standard Model gauge subgroup. One hallmark of these models is a natural hierarchy between gaugino and scalar masses. Another is the suppression of squark pair production due to vanishing amplitudes of certain processes, such as $q_{\text{L}} q_{\text{L}} \to \tilde{q}_{\text{L}} \tilde{q}_{\text{L}}$ via $t$-channel gluino \cite{Dudas:2014fr, Diessner:2017sq}. For these and other reasons, the parameter space of $R$-symmetric models remains far less constrained by searches for colored particles than that of the MSSM \cite{Kribs:2012ss, Alvarado:2018ch, Diessner:2019sq}.

The phenomenology of $R$-symmetric models is further enriched by the aforementioned adjoint fields, which form their own distinct sector with intricate phenomenology. This sector contains at least one complex scalar for each Standard Model gauge subgroup, all of which are $R$ neutral and able to decay to pairs of Standard Model particles. In the event that the components of these scalars are not degenerate and do not participate in any CP-violating interactions, it is convenient to view each complex field as a linear combination of a real scalar and a real pseudoscalar. The masses of these particles are minimally generated at the same time as the gaugino masses but can receive contributions from a wide variety of other SUSY-breaking operators. Such operators can generate considerable splitting between the masses of the scalar and pseudoscalar particles. The couplings of these particles to themselves and to the Dirac gauginos, the MSSM-like scalars, the Higgs fields, and the Standard Model particles generate a wealth of interesting dynamics.

In this work we explore the phenomenology of the $\mathrm{SU}(3)_{\text{c}}$ adjoint (color-octet) scalars --- the \emph{sgluons} \cite{Plehn:2008ae} --- in minimal $R$-symmetric models.  We map the parameter space for the scalar and pseudoscalar sgluons in light of the most recent data from the Large Hadron Collider (LHC). We provide a comprehensive catalog of sgluon decays, including nonstandard gauge-invariant diboson decays to gluon-photon ($g\gamma$) and gluon-$Z$ boson ($gZ$) and novel three-body decays. After considering the significant sgluon production modes, we discuss the phenomenology of these particles, identifying significant parameter space in which they are long-lived and decay with non-trivial displaced vertices. Using this information, we synthesize relevant LHC searches at $s^{1/2} = 7\, \text{TeV}$, $8\, \text{TeV}$, and $13\, \text{TeV}$ to extrapolate bounds on the parameter space of minimal $R$-symmetric models. We show that constraints on pair production of both scalar and pseudoscalar sgluons from searches for final states with four flavorless or heavy-flavor jets can be weakened by delayed decays and diminished branching fractions to Standard Model particles. Altogether we find significant regions where light color-octet scalars have not yet been ruled out.

This paper is organized as follows. In Sections \hyperref[s2]{2} and \hyperref[s3]{3}, we review minimal $R$-symmetric models, discussing the gauginos and adjoint scalars, describing the class of models considered in this work, and defining benchmark scenarios for our phenomenological investigation. In \hyperref[s4]{Section 4}, we systematically explore all relevant sgluon decays and compute both particles' production cross sections. In \hyperref[s5]{Section 5}, we describe the phenomenology of production and decay processes for the sgluons. We confront data from the LHC in \hyperref[s6]{Section 6} and discuss constraints on sgluon parameter space. Finally, in \hyperref[s7]{Section 7}, we draw conclusions and suggest future routes of inquiry.
\section{Review of \emph{R}-symmetric supersymmetry}
\label{s2}

In this section and the next we establish the background for our discussion of minimal $R$-symmetric models. We begin with a review of Dirac gauginos and adjoint scalars, describing the minimal generation of these particles' masses and their dynamics. We then broadly discuss $R$ symmetry, considering how to balance the need for exact or approximate symmetry with the desire for minimal particle content. In \hyperref[s3]{Section 3} we turn to the models that we study in the rest of this work, articulating our theoretical and phenomenological assumptions and defining several benchmarks for quantitative investigation.

\subsection{Dirac gauginos and adjoint scalars}\label{s2.A}

It is well known that Dirac gaugino masses can be minimally generated by a supersymmetry-breaking operator that only introduces finite radiative corrections. We begin with a review of this mechanism. Suppose that a hidden sector contains a $\text{U}(1)'$ gauge superfield $\mathcal{W}'$ with a nonvanishing $D$-term vacuum expectation value (VEV), denoted by $D'$, which breaks supersymmetry at a scale $\Lambda$. Then, in the infrared, the interactions between this $\text{U}(1)'$ superfield, the gauge superfields $\mathcal{W}_k$ for each Standard Model gauge subgroup $G_k$ (with $k \in \{1,2,3\}$ such that $G_3 = \mathrm{SU}(3)_{\text{c}}$), and the corresponding chiral adjoint superfields $\mathcal{A}_k$ are governed by the classic supersoft operators \cite{Fox:2002bu}
\begin{align}\label{e1}
\mathcal{L} \supset \sum_{k=1}^3 \int \d^2 \theta\, \frac{\kappa_k}{\Lambda}\, \mathcal{W}'^{\alpha} \mathcal{W}^a_{k\alpha} \mathcal{A}_k^a + \text{H.c.},
\end{align}
where summation is implied over $\alpha$ and $a$ (indices for, respectively, Weyl spinors and the adjoint representation of the gauge subgroup $G_k$). The dimensionless constants $\kappa_k$ parameterize the coupling of each adjoint superfield to each Standard Model gauge field. Integrating out the $\text{U}(1)'$ $D$ term yields Dirac gaugino masses, which we write with spinor indices suppressed as
\begin{align}\label{e2}
\mathcal{L} \supset -m_k(\lambda^a_k \psi^a_k + \text{H.c.}) \equiv -m_k\, \bar{\tilde{g}}^a_k \tilde{g}^a_k\ \ \ \text{with}\ \ \ m_k = \frac{\kappa_k}{\sqrt{2}}\frac{D'}{\Lambda}.
\end{align}
The couplings $\kappa_k$ can be unique, so the gaugino masses generated by the supersoft operator need not be unified. Note that \eqref{e2} fixes our notation for gauginos: the Dirac gaugino $\tilde{g}_k$ is created by the marriage of the Majorana gaugino $\lambda_k$ and the adjoint Majorana fermion $\psi_k$. By definition, unlike the Majorana gauginos, the Dirac gauginos are not self-charge-conjugate: $\tilde{g}_k^{\text{c}} \neq \tilde{g}_k$ \cite{Choi:2008gn}. A detailed exposition of the $\mathrm{SU}(3)_{\text{c}}$ \emph{Dirac gluinos} $\tilde{g}_3 \equiv \tilde{g}$, including an explicit definition of the charge-conjugated gluinos $\tilde{g}^{\text{c}}$, is provided in \hyperref[aA]{Appendix A}.

Next we consider the adjoint scalars. Here we restrict ourselves to the $\text{SU}(3)_{\text{c}}$ chiral adjoint (hence color-octet) scalar, the sgluon, but the following discussion applies in principle to all adjoint scalars. We decompose the complex color-octet scalar $\varphi_3$ according to
\begin{align}\label{e4}
\varphi_3^a \equiv \frac{1}{\sqrt{2}}(O^a + \ii o^a).
\end{align}
We assume for simplicity that the adjoint scalars do not violate CP, so that for instance $O$ is a scalar and $o$ a pseudoscalar. We denote the mass of the scalar sgluon by $m_O$ and the mass of the pseudoscalar by $m_o$. These masses, which are in general not equal, can receive contributions from multiple operators. An unavoidable mass contribution is generated by the supersoft operator \eqref{e1} and the canonical K\"{a}hler potential for the $\mathrm{SU}(3)_{\text{c}}$ adjoint superfield $\mathcal{A}_3$ and the superfields charged under the same gauge group. The interactions between sgluons and left-chiral squarks $\tilde{q}_{\text{L}}$, for instance, originate from
\begin{align}\label{e5}
    \mathcal{L} \supset \int \d^2 \theta\, \d^2 \theta^{\dagger} \left[\mathcal{Q}^{\dagger i} \exp \left\lbrace 2 g_3 [\bt{t}_3^c \mathcal{V}_3^c]_i^{\ j}\right\rbrace \mathcal{Q}_{j} + \mathcal{A}_3^{\dagger a}\exp \left\lbrace 2 g_3 [\bt{t}^c_3 \mathcal{V}^c_3]_a^{\ b}\right\rbrace \mathcal{A}_{3b}\right],
\end{align}
where $\mathcal{Q}$ is the superfield containing left-chiral quarks $q_{\text{L}}$ and squarks $\tilde{q}_{\text{L}}$. In this expression $\bt{t}_3$ are the generators of the appropriate representations of $\mathrm{SU}(3)$ (the difference is clear from both context and indices), and $g_3$ and $\mathcal{V}_3$ are respectively the $\mathrm{SU}(3)_{\text{c}}$ running coupling and vector superfield. The interactions and corresponding Feynman rules derived from \eqref{e5} are detailed in \hyperref[aA]{Appendix A}. When the $\mathrm{SU}(3)_{\text{c}}$ $D$ term is integrated out, the scalar receives a mass $m_O = 2m_3$ and the pseudoscalar remains massless. Additional supersymmetry-breaking operators also appear quite generally. For example, the scalar and pseudoscalar masses can be split by the \emph{lemon-twist} operators, which are also supersoft and cannot be forbidden by any symmetry that allows the operators \eqref{e1}:
\begin{align}\label{eL}
\mathcal{L} \supset \sum_{k=1}^3 \int \d^2 \theta\, \frac{\kappa'_k}{\Lambda^2}\, \mathcal{W}'^{\alpha} \mathcal{W}_{\alpha}' \mathcal{A}_k^a \mathcal{A}_k^a.
\end{align}
These operators give contributions to the squared mass of each adjoint that are large and positive for one component and negative for the other. This naturally raises concerns about a tachyonic mass for one component of the adjoint scalar. Various solutions to this problem have been proposed, most of which involve the inclusion of new operators. Such operators can either be postulated using symmetry arguments \cite{Carpenter:2010rsb}, or they can be generated via e.g. power expansions of $D$ term insertions in mass-generating diagrams, assuming messenger-based ultraviolet completions of the supersoft operator \cite{Carpenter:2015mna}. Given the variety of supersymmetry-breaking mass-generating operators, it is reasonable to make the phenomenological assumption that the masses of the components of the adjoints can be substantially split, and thus one component can justifiably be made light. Moreover, given the large number of possible ultraviolet completions, we can consider the scalar and pseudoscalar masses to be independent in a general infrared model.

\subsection{How to build a minimal $R$-symmetric model}\label{s2.B}

The supersoft operator \eqref{e1} possesses useful qualities beyond its avoidance of logarithmic divergences. In particular, it exhibits an $R$ symmetry, a symmetry with (at minimum) a $\mathrm{U}(1)$ subgroup that does not commute with supersymmetric transformations. An $R$ symmetry can forbid softly supersymmetry-breaking terms of the form
\begin{align}\label{e8}
 \mathcal{L} \supset -\frac{1}{2}\, M_k (\lambda_k^a\lambda_k^a + \text{H.c.})
 \end{align}
that give Majorana masses to (e.g., for $k=3$) the superpartner of the gluon. If such terms exist, the gauginos remain Majorana by definition. The Majorana gaugino and the adjoint fermion are defined to carry equal and opposite $R$ charge, so the marriage of the two remains a Dirac fermion as long as the $R$ symmetry is unbroken. For this reason alone, $R$ symmetry is theoretically well motivated, but the preservation of Dirac gauginos is far from the only potentially desirable consequence of $R$ symmetry. Others include the amelioration of the supersymmetric flavor problem, via exclusion of operators that simultaneously violate $R$ and flavor, and the elimination of mixing between left- and right-chiral squarks and sleptons \cite{Kribs:2009clsp, Chalons:2019md}. We frequently exploit the latter throughout this work.

Because there exists some freedom in the assignment of $R$ charge to other fields, there are a number of ways to endow a model with $R$ symmetry. It is common to conceive of a continuous $\mathrm{U}(1)_R$ symmetry as a generalization of discrete $R$ parity, so that all Standard Model particles are $R$ neutral. In this scheme, the $\mathrm{SU}(2)_{\text{L}}$-invariant contraction
\begin{align}\label{e9}
    \mathcal{L} \supset \int \d^2 \theta\, \mu\, \mathcal{H}_{\text{u}} \doot \mathcal{H}_{\text{d}}
\end{align}
of the up- and down-type chiral Higgs superfields $\mathcal{H}_{\text{u}}$ and $\mathcal{H}_{\text{d}}$ is forbidden. In models such as the Minimal $R$-Symmetric Supersymmetric Standard Model (MRSSM), this disaster is averted by adding new $R$-Higgs fields to generate Higgs masses while preserving $R$ neutrality among the MSSM-like scalar Higgs fields \cite{Kribs:2008rs}. While this approach successfully reinstates Higgsino (and Standard Model fermion) masses, it has drawbacks: the scalar sector of the MRSSM is significantly more complex than that of the MSSM, and an exact $R$ symmetry may be incompatible with quantum gravity \cite{Chalons:2019md}. However, there are alternatives that balance the desire for minimal particle content with the need to preserve the Dirac nature of the gauginos. In the Minimal Dirac Gaugino Supersymmetric Standard Model (MDGSSM), for example, the Higgs fields are allowed to have nonzero $R$ charge \cite{Benakli:2013mdg, Benakli:2014cmdg, Chalons:2019md}. An example set of $R$ charge assignments is displayed in \hyperref[tI]{Table 1}.
\begin{table}\label{tI}
\begin{center}
\begin{tabular}{| l || c | r || c | r || c | r |}
\hline
\rule{0pt}{3ex} & \ Superfield\ \ & $R$\ \ \ & Boson & $R$\ \ \ & \ Fermion\ \ & $R$\ \ \ \\[0.5ex]
\hline
\hline
\rule{0pt}{3.5ex}Gluon & $\mathcal{W}_3$ & \ +1\ \ & $g$ & 0\ \ & $\lambda_3$ & \ +1\\
\rule{0pt}{3.5ex}Left-chiral quark\ \ & $\mathcal{Q}$ & \ +1\ \ & $\tilde{q}_{\text{L}}$ & \ +1\ \ & $q_{\text{L}}$ & 0\\
\rule{0pt}{3.5ex}Right-chiral quark\ \ & $\overline{\mathcal{U}}{}^{\dagger},\overline{\mathcal{D}}{}^{\dagger}$ & \ 0\ \ & $\tilde{u}_{\text{R}}, \tilde{d}_{\text{R}}$ & \ 0\ \ & $u_{\text{R}},d_{\text{R}}$ & +1\\
\rule{0pt}{3.5ex}Higgs & $\ \mathcal{H_{\text{u}}},\mathcal{H}_{\text{d}}$\ \ & \ +1\ \ & \ $H_{\text{u}},H_{\text{d}}$\ \ & +1\ \ & \ $\tilde{H}_{\text{u}},\tilde{H}_{\text{d}}$\ \ & 0\\
\rule{0pt}{3.5ex}$\mathrm{SU}(3)_{\text{c}}$ adjoint\ \ & $\mathcal{A}_3$ & \ 0\ \ & $\varphi_3$ & 0\ \ & $\psi_3$ & $-$1\\[0.8ex]
\hline
\end{tabular}
\end{center}
\caption{$R$ charges of selected fields in a minimal model with $R$ symmetry broken only by a $B_{\mu}$ term.}
\end{table}
In this scheme, the $\mu$ term \eqref{e9} is $R$ symmetric, the gauginos remain Dirac, and the particle content of the scalar sector remains minimal. But if the Higgs fields carry $R$ charge, then electroweak symmetry breaking immediately induces spontaneous $R$ symmetry breaking, resulting in undesirable massless $R$-axions. A softly supersymmetry-breaking term of the form
\begin{align}\label{e10}
    \mathcal{L} \supset -B_{\mu}\,( H_{\text{u}} \doot H_{\text{d}} + \text{H.c.})
\end{align}
is therefore added to break $R$ uniquely and explicitly \cite{Chalons:2019md}. This system has two possible disadvantages: first, it suggests that supersymmetry should be broken by a different mechanism in the Higgs sector than in the rest of the model, since $\mu$ and $B_{\mu}$ should be of similar size; second, the explicit breaking of $R$ symmetry immediately imperils the Dirac gauginos. Neither of these issues are of concern for this work. First, there are already many gauge-mediated supersymmetry-breaking mechanisms that both preserve $R$ symmetry and generate mass spectra consistent with the models discussed above and throughout this work \cite{Nelson:1994rsb, Intriligator:2006rsb, Amigo:2009rsb, Carpenter:2010rsb}. Second, while explicit $R$ symmetry breaking does allow Majorana gaugino masses and left-right squark mixing, it has been comprehensively demonstrated that the properties of $R$-symmetric models in which we are presently interested remain mostly intact if the violation of $R$ symmetry remains small \cite{Choi:2008gn, Chalons:2019md}. We therefore assume that this condition is met. Finally, in these models, the sgluon can decay entirely to Standard Model particles via $R$-preserving operators. As such, $R$ symmetry breaking does not affect sgluon decay widths or production cross sections at tree level or one-loop level.

\section{Model parameters and benchmark scenarios}
\label{s3}

Altogether, then, we view minimal (approximately) $R$-symmetric models in the mold of the MDGSSM as well motivated, balancing minimality with attractive theoretical features and rich phenomenology. We investigate models of this class in the present work. In order to make this investigation concrete, we pause here to finish defining the Lagrangian that governs the sgluons, quarks, gluons, squarks, and Dirac gluinos with $R$ charges given by \hyperref[tI]{Table 1}. This Lagrangian is given compactly by
\begin{align}
    \mathcal{L} \supset \mathcal{L}_{\text{Dirac}} + \mathcal{L}_O + \mathcal{L}_{\text{soft}} + \int \d^2 \theta\, W_{\text{MSSM}}.
\end{align}
In this expression, $\mathcal{L}_{\text{Dirac}}$ comprises the masses of the Dirac bino $\tilde{B}$, wino $\tilde{W}$, and gluino $\tilde{g}$,
\begin{align}\label{eDirac}
\mathcal{L}_{\text{Dirac}} = - \left[m_1 \bar{\tilde{B}}\tilde{B} + m_2\, \overbar{\tilde{W}}\tilde{W} + m_3\, \bar{\tilde{g}}\tilde{g}\right],
\end{align}
which are generated by \eqref{e1}; $\mathcal{L}_O$ contains the sgluon kinetic and gauge terms \eqref{eA1}, which are generated by the supersoft operator \eqref{e1} and the K\"{a}hler potential (\eqref{e5} and the like); $\mathcal{L}_{\text{soft}}$ includes the softly supersymmetry-breaking terms,
\begin{multline}\label{Lagrangian}
    \mathcal{L}_{\text{soft}} \supset -\bigg[ m_O^2|O|^2 + m_o^2 |o|^2 + m_{Q_i}^2 |\tilde{q}_{\text{L}i}|^2+m_{u_i}^2 |\tilde{u}_{\text{R}i}|^2+m_{d_i}^2 |\tilde{d}_{\text{R}i}|^2\\+ m_{H_{\text{u}}}^2|H_{\text{u}}|^2 + m_{H_{\text{d}}}^2|H_{\text{d}}|^2+ B_{\mu}\,( H_{\text{u}} \doot H_{\text{d}} + \text{H.c.}) \bigg],
\end{multline}
where, just for the moment, $i$ indexes not color but generation; and, finally, $W_{\text{MSSM}}$ is the MSSM superpotential, given in part by
\begin{align}\label{Wterms}
    W_{\text{MSSM}} \supset \mu\, \mathcal{H}_{\text{u}} \doot \mathcal{H}_{\text{d}} + \overline{\mathcal{U}} \bt{Y}_{\!\!\text{u}} \mathcal{Q} \doot \mathcal{H}_{\text{u}} - \overline{\mathcal{D}} \bt{Y}_{\!\!\text{d}} \mathcal{Q} \doot \mathcal{H}_{\text{d}}.
\end{align}

The relevant parameters in the strong sector are the masses of the Dirac gluino, the sgluons, and the squarks. These, along with the electroweak gaugino parameters ($\mu$ and the masses $m_1$ and $m_2$ of the Dirac bino and wino) fully determine the mass spectrum of the particles we study in this work. In order to facilitate this study, we define several benchmark scenarios with a variety of spectra that highlight the most interesting features of minimal $R$-symmetric models. Below we broadly characterize these spectra and summarize the assumptions, many of which follow from the discussion in \hyperref[s2]{Section 2}, that inform our choices.

\begin{enumerate}
\item As we hinted above, we take a phenomenological approach throughout and assume that the masses of the squarks, the gluino, and the sgluons can be varied independently. For example, we examine regions of parameter space where the sgluons are lighter than the squarks and the gluino. Accessing such regions in parameter space might be realized with large supersoft contributions to the squarks from heavy bino and wino masses \cite{Carpenter:2017xru} or in more elaborate theoretical scenarios. This phenomenological approach requires an ultraviolet completion to supply the full set of operators contributing to relevant soft mass parameters. However, even in this case, the particle content and parameter space of the infrared theory can be viewed as minimal. 

\item We assume that the Higgs sector provides the only source of $R$ symmetry breaking. As such, we suppress $R$-breaking trilinear ($A$) terms for the MSSM-like scalars. This restriction eliminates mixing at tree level between left- and right-chiral squarks, so that e.g. the top squark (stop) mass eigenstates are $\tilde{t}_{\text{L}} \equiv \tilde{t}_1$ and $\tilde{t}_{\text{R}} \equiv \tilde{t}_2$. We take these to be either reasonably light ($\sim\! 1\, \text{TeV}$) or moderately heavy (multi-TeV). The squarks of all other flavors (e.g. the left- and right-chiral bottom squarks $\tilde{b}_{\text{L}}, \tilde{b}_{\text{R}}$) are degenerate and decoupled.

\item Also in the interest of preserving $R$ symmetry, we omit trilinear terms from the superpotential and $A$-like trilinear terms from the scalar potential, which couple the new adjoint scalars to each other.  

\item The choice of LSP has a large effect on decay chains and collider limits on superpartners in Dirac gaugino models \cite{Carpenter:2016lgo}.  We choose to study a conservative LSP scenario natural to models with heavy Dirac gaugino masses. Namely, we assume that the electroweakinos --- also (approximately) Dirac --- display a natural hierarchy such that the lightest and next-to-lightest supersymmetric particles (LSP $\tilde{\chi}_1^0$ and NLSP $\tilde{\chi}_1^{\pm}$) are  Higgsino-like, with masses given at tree level by $\mu$ \cite{Choi:2010dn, Kribs:2009clsp}. $\mu$ is taken to be small enough compared to the Dirac masses $m_k$ to avoid spoiling the other qualitative features of $R$-symmetric models.
\end{enumerate}

The benchmarks we have chosen following the above considerations are displayed in \hyperref[tII]{Table 2}.
\begin{table}\label{tII}
\begin{center}
\begin{tabular}{| r || c | c | c | c | c | c |}
\hline
\rule{0pt}{3ex} & B1 & B2 & B3 & B4 & B5 & B6\\[0.5ex]
\hline
\hline
\rule{0pt}{3.5ex}$m_{\tilde{t}_1}\, \text{(GeV)}$\ \ & \ 1000\ \ & \ 1000\ \ & \ 1500\ \ & \ 1500\ \ & \ 800\ \ & \ 2000\ \ \\
\rule{0pt}{3.5ex}$m_{\tilde{t}_2}\, \text{(GeV)}$\ \ & 1500 & 1500 & 2000 & 2000 & 900 & 2500\\
\rule{0pt}{3.5ex}$m_3\, \text{(GeV)}$\ \ & 5000 & 3500 & 5000 & 3500 & 3000 & 2000\\
\rule{0pt}{3.5ex}$\mu\, \text{(GeV)}$\ \ & 300 & 300 & 500 & 500 & 250 & 300\\[0.8ex]
\hline
\end{tabular}
\end{center}
\caption{Benchmarks for quantitative investigation of sgluon decay and production.}
\end{table}
The fifth and sixth benchmarks, B5 and B6, are less conservative than the first four. B5 closely approaches even the relaxed $R$-symmetric stop and gluino mass limits \cite{Diessner:2019sq}, and B6 is a unique benchmark in which we abandon the natural hierarchy between Dirac gauginos and scalars. Unmentioned in \hyperref[tII]{Table 2} are scenarios where we probe the mass splitting of the stops. This is of interest because the stop-mediated one-loop decays of sgluons to quarks and gluons vanish if the stops are degenerate. We make clear in Sections \hyperref[s5]{5} and \hyperref[s6]{6} wherever we depart from the benchmarks displayed here.
\section{Decays and production of color-octet scalars}
\label{s4}

In this section we present the analytic results that form the basis of our phenomenological investigation of color-octet scalars in Sections \hyperref[s5]{5} and \hyperref[s6]{6}. We first review the significant decay channels of scalar and pseudoscalar sgluons and compute the associated partial decay rates. We then compute the cross sections of single production of the scalar sgluon and of pair production of both particles.

\subsection{Partial rates of decay}\label{s4.A}

\begin{figure}\label{f1}
\begin{align*}
&(\text{a})\ \ \ \ \ \scalebox{0.75}{\begin{tikzpicture}[baseline={([yshift=-.5ex]current bounding box.center)},xshift=12cm]
\begin{feynman}[large]
\vertex (i1);
\vertex [right = 1.5cm of i1] (i2);
\vertex [above right=1.5 cm of i2] (v1);
\vertex [below right=1.5cm of i2] (v2);
\diagram* {
(i1) -- [ultra thick, scalar] (i2),
(v2) -- [ultra thick, charged scalar] (i2),
(i2) -- [ultra thick, charged scalar] (v1),
};
\end{feynman}
\node at (2.75,0.75) {$\tilde{q}_{\text{L}}$};
\node at (2.75,-0.7) {$\tilde{q}^{\dagger}_{\text{L}}$};
\node at (0.2,0.3) {$O$};
\end{tikzpicture}}\ \ \ \ \ \text{and}\ \ \ \ \ \scalebox{0.75}{\begin{tikzpicture}[baseline={([yshift=-.5ex]current bounding box.center)},xshift=12cm]
\begin{feynman}[large]
\vertex (i1);
\vertex [right = 1.5cm of i1] (i2);
\vertex [above right=1.5 cm of i2] (v1);
\vertex [below right=1.5cm of i2] (v2);
\diagram* {
(i1) -- [ultra thick, scalar] (i2),
(v2) -- [ultra thick, charged scalar] (i2),
(i2) -- [ultra thick, charged scalar] (v1),
};
\end{feynman}
\node at (2.75,0.75) {$\tilde{q}_{\text{R}}$};
\node at (2.75,-0.7) {$\tilde{q}^{\dagger}_{\text{R}}$};
\node at (0.2,0.3) {$O$};
\end{tikzpicture}}\\[3.5ex]
&(\text{b})\ \ \ \ \ \scalebox{0.75}{\begin{tikzpicture}[baseline={([yshift=-.5ex]current bounding box.center)},xshift=12cm]
\begin{feynman}[large]
\vertex (i1);
\vertex [right = 1.5cm of i1] (i2);
\vertex [above right=1.5 cm of i2] (v1);
\vertex [below right=1.5cm of i2] (v2);
\diagram* {
(i1) -- [ultra thick, scalar] (i2),
(v2) -- [ultra thick, fermion] (i2),
(i2) -- [ultra thick, photon] (v2),
(i2) -- [ultra thick, fermion] (v1),
(i2) -- [ultra thick, photon] (v1),
};
\end{feynman}
\node at (2.75,0.75) {$\tilde{g}$};
\node at (2.75,-0.7) {$\bar{\tilde{g}}$};
\node at (0.2,0.3) {$O$};
\node at (0.6,-0.3) {\text{or}\, $o$};
\end{tikzpicture}}
\end{align*}
\caption{Diagrams for (a) scalar sgluon decays to squarks and (b) scalar or pseudoscalar decays to gluinos.}
\end{figure}
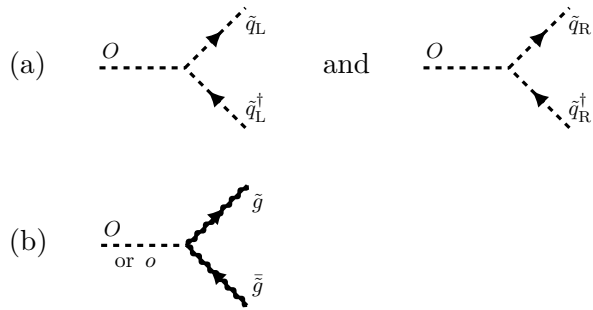

The sgluons can decay at tree level and one-loop level through a variety of channels, though not all channels are available to both particles. We begin with the simplest two-body tree-level decays. The scalar sgluon can decay at this level to pairs of left- or right-chiral squarks $\tilde{q}^{\dagger}\tilde{q}$. These channels are closed to the pseudoscalar: only the scalar sgluon couples directly to pairs of MSSM-like scalars. On the other hand, both sgluons can decay at tree level to gluino pairs $\bar{\tilde{g}}\tilde{g}$. The operators that allow these decays are given explicitly by \eqref{eA1}. The diagrams for these decays are provided in \hyperref[f1]{Figure 1}. These and subsequent diagrams were generated using the \LaTeX\ package \textsc{Tikz-Feynman} \cite{Ellis:2017fd}.
The rates of these decays are given by \cite{Choi:2009co}
\begin{align}\label{e11}
\nonumber \Gamma(O \to \tilde{q}^{\dagger}\tilde{q}) &= \frac{1}{2}\, \alpha_3\, \frac{m_3^2}{m_O}\, \beta_{\tilde{q}},\\
\nonumber \Gamma(O \to \bar{\tilde{g}}\tilde{g}) &= \frac{3}{2}\, \alpha_3\, m_O\, \beta_{\tilde{g}}^3,\\
\text{and}\ \ \ \Gamma(o \to \bar{\tilde{g}}\tilde{g}) &= \frac{3}{2}\, \alpha_3\, m_o\, \beta_{\tilde{g}},
\end{align}
where as an estimate $4\pi \alpha_3= g_3^2$ is renormalization-group (RG) evolved to the sgluon mass from $\alpha_3(m_Z^2) \approx 0.12$ using the MSSM one-loop $\mathrm{SU}(3)_{\text{c}}$ $\beta$-function \cite{Martin:1997sp}, and where the kinematic function $\beta_X=[1 - 4(m_X m_O^{-1})^2]^{1/2}$ is the speed of the final-state particles in a decay to an $X$ pair.

Due to their $R$ neutrality (viz. \hyperref[tI]{Table 1}), the scalar and pseudoscalar sgluons can decay to pairs of Standard Model particles. In particular, decays to pairs of gauge bosons and quarks appear at one-loop level, though only the decays to quarks are open to the pseudoscalar. The diagrams for decays to gluons and quarks are provided in \hyperref[f2]{Figure 2}, with colors and momenta labeled to facilitate the discussion in \hyperref[aB]{Appendix B}.
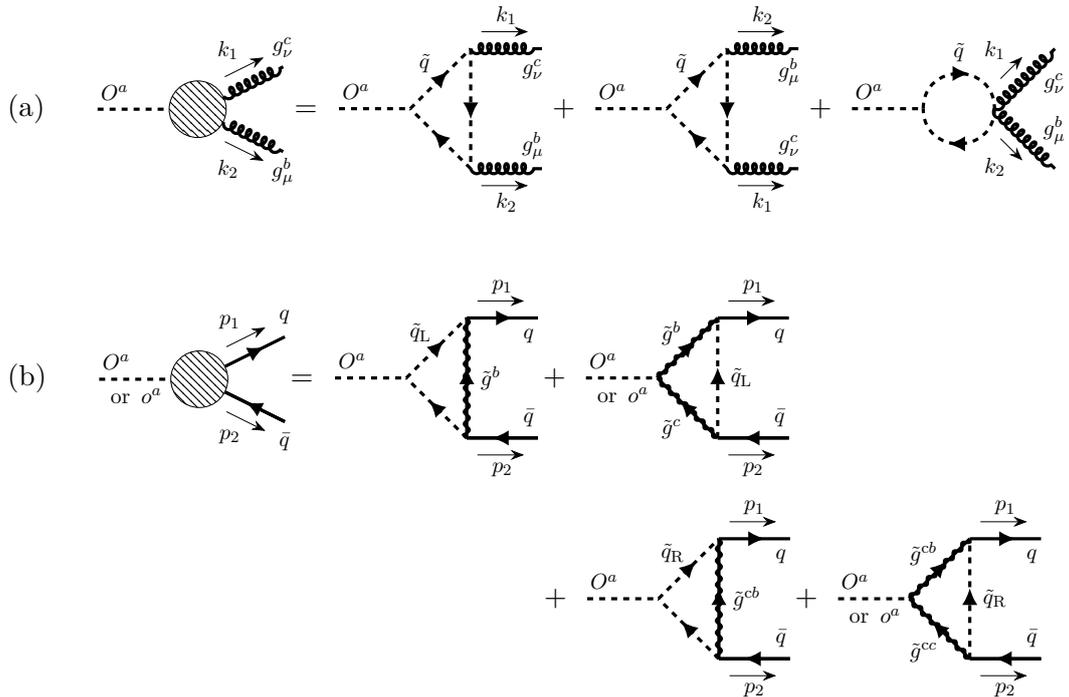
\begin{figure}\label{f2}
\begin{align*}
&(\text{a})\ \ \ \ \ \scalebox{0.75}{\begin{tikzpicture}[baseline={([yshift=-.5ex]current bounding box.center)},xshift=12cm]
\begin{feynman}[large]
\vertex (i1);
\vertex [right = 1.25cm of i1, blob] (i2){};
\vertex [above right=0.75 cm and 1.5cm of i2] (v1);
\vertex [below right=0.75cm and 1.5cm of i2] (v2);
\diagram* {
(i1) -- [ultra thick, scalar] (i2),
(i2) -- [ultra thick, gluon,momentum=$k_1$] (v1),
(i2) -- [ultra thick, gluon,momentum'=$k_2$] (v2),
};
\end{feynman}
\node at (0.3,0.3) {$O^a$};
\node at (3.25,1.1) {$g^c_{\nu}$};
\node at (3.25,-1.1) {$g^b_{\mu}$};
\end{tikzpicture}}\! \! \! =\ \scalebox{0.75}{\begin{tikzpicture}[baseline={([yshift=-0.75ex]current bounding box.center)},xshift=12cm]
\begin{feynman}[large]
\vertex (i1);
\vertex [right = 1.25cm of i1] (i2);
\vertex [above right=1.5 cm of i2] (v1);
\vertex [below right=1.5cm of i2] (v2);
\vertex [right=1.25cm of v1] (g1);
\vertex [right=1.25cm of v2] (g2);
\diagram* {
(i1) -- [ultra thick, scalar] (i2),
(i2) -- [ultra thick, charged scalar] (v1),
(v1) -- [ultra thick, charged scalar] (v2),
(v2) -- [ultra thick, charged scalar] (i2),
(v1) -- [ultra thick, gluon, momentum=$k_1$] (g1),
(v2) -- [ultra thick, gluon, momentum'=$k_2$] (g2),
};
\end{feynman}
\node at (0.3,0.3) {$O^a$};
\node at (3.4,0.7) {$g^c_{\nu}$};
\node at (3.4,-0.65) {$g^b_{\mu}$};
\node at (1.5,0.8) {$\tilde{q}$};
\end{tikzpicture}}\! + \ \scalebox{0.75}{\begin{tikzpicture}[baseline={([yshift=-0.75ex]current bounding box.center)},xshift=12cm]
\begin{feynman}[large]
\vertex (i1);
\vertex [right = 1.25cm of i1] (i2);
\vertex [above right=1.5 cm of i2] (v1);
\vertex [below right=1.5cm of i2] (v2);
\vertex [right=1.25cm of v1] (g1);
\vertex [right=1.25cm of v2] (g2);
\diagram* {
(i1) -- [ultra thick, scalar] (i2),
(i2) -- [ultra thick, charged scalar] (v1),
(v1) -- [ultra thick, charged scalar] (v2),
(v2) -- [ultra thick, charged scalar] (i2),
(v1) -- [ultra thick, gluon, momentum=$k_2$] (g1),
(v2) -- [ultra thick, gluon, momentum'=$k_1$] (g2),
};
\end{feynman}
\node at (0.3,0.3) {$O^a$};
\node at (3.4,0.65) {$g^b_{\mu}$};
\node at (3.4,-0.65) {$g^c_{\nu}$};
\node at (1.5,0.8) {$\tilde{q}$};
\end{tikzpicture}}\! +\ \scalebox{0.75}{\begin{tikzpicture}[baseline={([yshift=-.75ex]current bounding box.center)},xshift=12cm]
\begin{feynman}[large]
\vertex (i1);
\vertex [right = 1.25cm of i1] (i2);
\vertex [right= 1.25cm of i2] (g1);
\vertex [above right=1.5 cm of g1] (v1);
\vertex [below right=1.5cm of g1] (v2);
\diagram* {
(i1) -- [ultra thick, scalar] (i2),
(i2) -- [ultra thick, charged scalar, half left, looseness=1.7] (g1),
(g1) -- [ultra thick, charged scalar, half left, looseness=1.7] (i2),
(g1) -- [ultra thick, gluon,momentum={[arrow shorten=0.3]$k_1$}] (v1),
(g1) -- [ultra thick, gluon,momentum'={[arrow shorten=0.3]$k_2$}] (v2),
};
\end{feynman}
\node at (0.3,0.3) {$O^a$};
\node at (3.55,0.45) {$g^c_{\nu}$};
\node at (3.55,-0.375) {$g^b_{\mu}$};
\node at (1.875,1) {$\tilde{q}$};
\end{tikzpicture}}\\[3.5ex]
&(\text{b})\ \ \ \ \ \begin{multlined}[t][12.49cm]\scalebox{0.75}{\begin{tikzpicture}[baseline={([yshift=-.5ex]current bounding box.center)},xshift=12cm]
\begin{feynman}[large]
\vertex (i1);
\vertex [right = 1.25cm of i1, blob] (i2){};
\vertex [above right=0.75 cm and 1.5cm of i2] (v1);
\vertex [below right=0.75cm and 1.5cm of i2] (v2);
\diagram* {
(i1) -- [ultra thick, scalar] (i2),
(i2) -- [ultra thick, fermion,momentum=$p_1$] (v1),
(i2) -- [ultra thick ,momentum'=$p_2$] (v2),
(v2) -- [ultra thick, fermion] (i2),
};
\end{feynman}
\node at (0.3,0.3) {$O^a$};
\node at (0.65,-0.3) {\text{or}\, $o^a$};
\node at (3.25,1.1) {$q$};
\node at (3.25,-1.1) {$\bar{q}$};
\end{tikzpicture}}\! \! \! =\ \scalebox{0.75}{\begin{tikzpicture}[baseline={([yshift=-0.75ex]current bounding box.center)},xshift=12cm]
\begin{feynman}[large]
\vertex (i1);
\vertex [right = 1.25cm of i1] (i2);
\vertex [above right=1.5 cm of i2] (v1);
\vertex [below right=1.5cm of i2] (v2);
\vertex [right=1.25cm of v1] (g1);
\vertex [right=1.25cm of v2] (g2);
\diagram* {
(i1) -- [ultra thick, scalar] (i2),
(i2) -- [ultra thick, charged scalar] (v1),
(v1) -- [ultra thick, photon] (v2),
(v2) -- [ultra thick, fermion] (v1),
(v2) -- [ultra thick, charged scalar] (i2),
(v1) -- [ultra thick, fermion, momentum=$p_1$] (g1),
(v2) -- [ultra thick, momentum'=$p_2$] (g2),
(g2) -- [ultra thick, fermion] (v2),
};
\end{feynman}
\node at (0.3,0.3) {$O^a$};
\node at (3.4,0.75) {$q$};
\node at (3.4,-0.70) {$\bar{q}$};
\node at (1.5,0.8) {$\tilde{q}_{\text{L}}$};
\node at (2.73,0) {$\tilde{g}^b$};
\end{tikzpicture}}\! + \ \scalebox{0.75}{\begin{tikzpicture}[baseline={([yshift=-0.75ex]current bounding box.center)},xshift=12cm]
\begin{feynman}[large]
\vertex (i1);
\vertex [right = 1.25cm of i1] (i2);
\vertex [above right=1.5 cm of i2] (v1);
\vertex [below right=1.5cm of i2] (v2);
\vertex [right=1.25cm of v1] (g1);
\vertex [right=1.25cm of v2] (g2);
\diagram* {
(i1) -- [ultra thick, scalar] (i2),
(i2) -- [ultra thick, fermion] (v1),
(i2) -- [ultra thick, photon] (v1),
(v2) -- [ultra thick, charged scalar] (v1),
(v2) -- [ultra thick, fermion] (i2),
(v2) -- [ultra thick, photon] (i2),
(v1) -- [ultra thick, fermion, momentum=$p_1$] (g1),
(v2) -- [ultra thick, momentum'=$p_2$] (g2),
(g2) -- [ultra thick, fermion] (v2),
};
\end{feynman}
\node at (0.3,0.3) {$O^a$};
\node at (0.65,-0.3) {\text{or}\, $o^a$};
\node at (3.4,0.75) {$q$};
\node at (3.4,-0.70) {$\bar{q}$};
\node at (1.5,0.8) {$\tilde{g}^b$};
\node at (1.5,-0.9) {$\tilde{g}^c$};
\node at (2.73,0) {$\tilde{q}_{\text{L}}$};
\end{tikzpicture}}\\ + \ \scalebox{0.75}{\begin{tikzpicture}[baseline={([yshift=-0.75ex]current bounding box.center)},xshift=12cm]
\begin{feynman}[large]
\vertex (i1);
\vertex [right = 1.25cm of i1] (i2);
\vertex [above right=1.5 cm of i2] (v1);
\vertex [below right=1.5cm of i2] (v2);
\vertex [right=1.25cm of v1] (g1);
\vertex [right=1.25cm of v2] (g2);
\diagram* {
(i1) -- [ultra thick, scalar] (i2),
(i2) -- [ultra thick, charged scalar] (v1),
(v1) -- [ultra thick, photon] (v2),
(v2) -- [ultra thick, fermion] (v1),
(v2) -- [ultra thick, charged scalar] (i2),
(v1) -- [ultra thick, fermion, momentum=$p_1$] (g1),
(v2) -- [ultra thick, momentum'=$p_2$] (g2),
(g2) -- [ultra thick, fermion] (v2),
};
\end{feynman}
\node at (0.3,0.3) {$O^a$};
\node at (3.4,0.75) {$q$};
\node at (3.4,-0.70) {$\bar{q}$};
\node at (1.5,0.8) {$\tilde{q}_{\text{R}}$};
\node at (2.8,0) {$\tilde{g}^{\text{c}b}$};
\end{tikzpicture}}\! + \ \scalebox{0.75}{\begin{tikzpicture}[baseline={([yshift=-0.75ex]current bounding box.center)},xshift=12cm]
\begin{feynman}[large]
\vertex (i1);
\vertex [right = 1.25cm of i1] (i2);
\vertex [above right=1.5 cm of i2] (v1);
\vertex [below right=1.5cm of i2] (v2);
\vertex [right=1.25cm of v1] (g1);
\vertex [right=1.25cm of v2] (g2);
\diagram* {
(i1) -- [ultra thick, scalar] (i2),
(i2) -- [ultra thick, fermion] (v1),
(i2) -- [ultra thick, photon] (v1),
(v2) -- [ultra thick, charged scalar] (v1),
(v2) -- [ultra thick, fermion] (i2),
(v2) -- [ultra thick, photon] (i2),
(v1) -- [ultra thick, fermion, momentum=$p_1$] (g1),
(v2) -- [ultra thick, momentum'=$p_2$] (g2),
(g2) -- [ultra thick, fermion] (v2),
};
\end{feynman}
\node at (0.3,0.3) {$O^a$};
\node at (0.65,-0.3) {\text{or}\, $o^a$};
\node at (3.4,0.75) {$q$};
\node at (3.4,-0.70) {$\bar{q}$};
\node at (1.5,0.8) {$\tilde{g}^{\text{c}b}$};
\node at (1.5,-0.95) {$\tilde{g}^{\text{c}c}$};
\node at (2.75,0) {$\tilde{q}_{\text{R}}$};
\end{tikzpicture}}
\end{multlined}
\end{align*}
\caption{Diagrams for (a) scalar sgluon decays to gluons and (b) scalar or pseudoscalar decays to quark-antiquark pairs. Either $\tilde{q}_{\text{L}}$ or $\tilde{q}_{\text{R}}$ can run in the $O \to gg$ loops. $\tilde{q}_{\text{R}}$ run with charge-conjugated gluinos $\tilde{g}^{\text{c}}$ in the $O\, \text{or}\, o \to \bar{q}q$ loops. Not all diagrams contribute to the pseudoscalar decay.}
\end{figure}
The rate of decay of the scalar to a gluon pair $gg$ is given by \cite{Choi:2009co}
\begin{align}\label{e12}
\Gamma(O\to gg) = \frac{5}{192\pi^2}\, \alpha_3^3\, \frac{m_3^2}{m_O} \left|\mathcal{F}(O \to gg)\right|^2,
\end{align}
where the form factor $\mathcal{F}(O \to gg)$ is given by \eqref{eB1}. This decay is highly sensitive, via $\mathcal{F}(O \to gg)$, to the mass difference between left- and right-chiral squarks: in particular, the rates vanish if the squarks are mass degenerate. Since we take all squarks except stops to be degenerate, $\mathcal{F}(O \to gg)$ only receives contributions from stops. 

The decay of a colored resonance to a jet and a photon was explored in a model-independent way some time ago \cite{Englert:2017gluphot}. This signature is potentially interesting, especially for pair-produced colored resonances, because final states with one or two photons could be more sensitive than multijet final states, to some extent regardless of the specific nature of the resonance \cite{Cacciapaglia:2020gluphot}. This motivates us to search for sgluon decays involving an electroweak gauge boson. We find that the scalar sgluon in minimal $R$-symmetric models is perfectly capable of decaying to $g\gamma$ in a manner highly analogous to the $gg$ decay. The rate of this decay --- if mediated only by stops --- can be written as
\begin{align}\label{eB}
\Gamma(O \to g \gamma) = \frac{8}{15}\frac{\alpha_1}{\alpha_3}\, \Gamma(O\to gg) \cos^2 \theta_{\text{w}},
\end{align}
where $4\pi \alpha_1 = g_1^2$ is the weak hypercharge coupling and $\theta_{\text{w}}$ is the weak mixing angle, both estimated to RG evolve as in the MSSM. The ratio of $\Gamma(O \to gg)$ to $\Gamma(O \to g\gamma)$ is fixed by gauge invariance. If all evolving observables are evaluated at $m_Z$, the rate of decay to gluons is a bit less than thirty times larger than the rate of decay to $g\gamma$. Before we move on, we note that the scalar also naturally decays to $gZ$ at roughly a third of the rate of decay to $g\gamma$.

Both sgluons can decay to quark-antiquark pairs $\bar{q}q$ at one-loop level. The rates of these decays are given by \cite{Goodsell:2014dg}
\begin{align}\label{e13}
\nonumber \Gamma(O \to \bar{q}q) &= \frac{9}{64\pi^2}\, \alpha_3^3\, m_O(m_3 m_q)^2\, \beta_q^3 \left|\mathcal{F}(O \to \bar{q}q)\right|^2\\
\text{and}\ \ \ \Gamma(o \to \bar{q}q) &= \frac{9}{64\pi^2}\, \alpha_3^3\, m_o(m_3m_q)^2\, \beta_q \left|\mathcal{F}(o \to \bar{q}q)\right|^2,
\end{align}
where the form factors $\mathcal{F}(O \to \bar{q}q)$ and $\mathcal{F}(o \to \bar{q}q)$ are given by \eqref{eB7} and \eqref{eB10}. Much like $\Gamma(O \to gg)$, these decay rates vanish if the left- and right-chiral squarks are degenerate. They are also proportional to the mass of the external quarks, so only the decays to third-generation quarks are considerable, with the $\bar{t}t$ decay dominating whenever it is kinematically accessible. When it is not, the $\bar{b}b$ decay takes over. This is particularly important for the pseudoscalar, which has no other available decays below $m_o = 2m_t$.

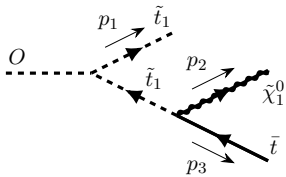
\begin{figure}\label{f3}
\begin{align*}
\scalebox{0.75}{\begin{tikzpicture}[baseline={([yshift=1.25ex]current bounding box.center)},xshift=12cm]
\begin{feynman}[large]
\vertex (i1);
\vertex [right=1.5 cm of i1] (g1);
\vertex [above right=0.75cm and 1.5cm of g1] (f1);
\vertex [below right = 0.75 cm and 1.5cm of g1] (i2);
\vertex [above right=0.8 cm and 1.6cm of i2] (v1);
\vertex [below right=0.8cm and 1.6cm of i2] (v2);
\diagram* {
(i1) -- [ultra thick, scalar] (g1),
(g1) -- [ultra thick, charged scalar, momentum={[arrow shorten=0.25]$p_1$}] (f1),
(i2) -- [ultra thick, charged scalar] (g1),
(i2) -- [ultra thick, photon,momentum={[arrow shorten=0.25]$p_2$}] (v1),
(i2) -- [ultra thick, fermion] (v1),
(i2) -- [ultra thick, momentum'={[arrow shorten=0.25]$p_3$}] (v2),
(v2) -- [ultra thick, fermion] (i2),
};
\end{feynman}
\node at (0.2,0.3) {$O$};
\node at (2.75,1) {$\tilde{t}_{1}$};
\node at (2.6,-0.1) {$\tilde{t}_1$};
\node at (4.7,-0.35) {$\tilde{\chi}_1^0$};
\node at (4.7,-1.275) {$\bar{t}$};
\end{tikzpicture}}
\end{align*}
\caption{Diagram for scalar sgluon decay to a light stop, a top antiquark, and the lightest neutralino, assumed to be pure Higgsino. The other significant $\tilde{t}_1$-mediated decay is given by replacing $\bar{t} \to \bar{b}$ and $\tilde{\chi}_1^0 \to \tilde{\chi}_1^-$. Similar decays exist for heavy stops.}
\end{figure}

Finally, the scalar can undergo three-body decays at tree level to a squark, the LSP or NLSP, and the corresponding quark. In the simple models under consideration, where the LSP $\tilde{\chi}_1^0$ and the NLSP $\tilde{\chi}_1^{\pm}$ are Dirac Higgsinos, the rates of these decays are proportional to the Yukawa coupling of the involved quark --- viz. \hyperref[aA]{Appendix A} --- and are thus dominated by the third-generation (s)quark channels even if other channels are kinematically allowed. In models with significant electroweakino mixing, these three-body decays also include light-flavor squarks, but the results are similar. The diagram for the decay to a light stop, a top antiquark, and the lightest neutralino is provided in \hyperref[f3]{Figure 3}, with momenta labeled to clarify phase-space integration.
The rate of this decay is given by
\begin{align}\label{e14}
\Gamma(O \to \tilde{t}_1 \bar{t}\tilde{\chi}_1^0) = \frac{1}{32\pi}\, \alpha_3\, \frac{m_3^2}{m_O}\, y_t^2 \int_{(m_t+m_{\tilde{\chi}_1^0})^2}^{(m_O - m_{\tilde{t}_1})^2} \d M^2\, \mathcal{F}(O \to \tilde{t}_1 \bar{t}\tilde{\chi}_1^0),
\end{align}
where the form factor $\mathcal{F}(O \to \tilde{t}_1 \bar{t}\tilde{\chi}_1^0)$ is given by \eqref{eB12} as a function of the subsystem invariant $M^2 = (p_2+p_3)^2$. The decay to a light stop, a bottom antiquark, and the lightest chargino takes the same form with appropriate mass substitutions in the phase-space integral. In order to tame singular behavior at the thresholds for decays to on-shell squark pairs, we include the decay widths of the stops, assuming that $99\%$ of their widths are given by decays to $\bar{t}\tilde{\chi}_1^0$ and $\bar{b}\tilde{\chi}_1^-$. Taken together, these decays are competitive with the loop-induced decays near the threshold for scalar decays to $\tilde{t}_1$ pairs.

\subsection{Production cross sections}\label{s4.B}

We now turn to sgluon production, which like the decays can proceed via a number of channels. Only the scalar can be singly produced, but both particles can be produced in pairs. We begin with the former production mode. At the LHC, single production of the scalar is due almost entirely to gluon fusion.\footnote{This is certainly true in any scenario where the loop-induced coupling of the scalar to gluons dwarfs that to quark-antiquark pairs. But this result is actually generic since the latter coupling is proportional to the quark mass, and this creates tension between the coupling strength and the associated parton distribution functions.} This production process is closely related to the decay of the scalar to two gluons: the diagrams for production are simply given by \hyperref[f2]{Figure 2(a)} with momenta and the flow of time reversed, so the amplitudes for these processes are identical. The hadron-level cross section of this production mode is given in terms of the corresponding decay rate \eqref{e12} by
\begin{align}\label{e15}
\sigma(gg \to O) = \frac{\pi^2}{m_O}\, \Gamma(O \to gg)\, \frac{1}{s} \int_{m_O^2/s}^1 \d x\, \frac{1}{x}\, g(x,m_O^2) g(m_O^2/sx,m_O^2),
\end{align}
where $s$ is the squared center-of-mass energy and $g(x,q^2)$ is the gluon distribution function with momentum fraction $x$ at factorization scale $q$. For simplicity we take the renormalization and factorization scales to coincide at the mass of the daughter sgluon.

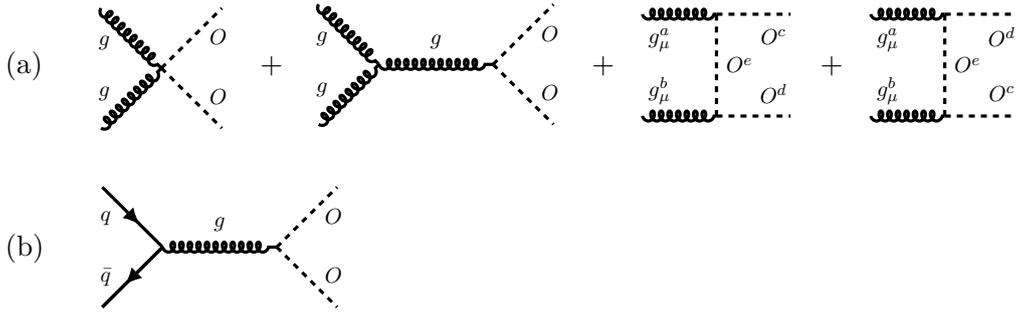
\begin{figure}\label{f4}
\begin{align*}
&(\text{a})\ \ \ \ \ \scalebox{0.75}{\begin{tikzpicture}[baseline={([yshift=-.5ex]current bounding box.center)},xshift=12cm]
\begin{feynman}[large]
\vertex (i1);
\vertex [above left = 1.5cm of i1] (g1);
\vertex [below left = 1.5cm of i1] (g2);
\vertex [above right=1.5 cm of i1] (v1);
\vertex [below right=1.5cm of i1] (v2);
\diagram* {
(g1) -- [ultra thick, gluon] (i1),
(g2) -- [ultra thick, gluon] (i1),
(i1) -- [ultra thick, scalar] (v1),
(i1) -- [ultra thick, scalar] (v2),
};
\end{feynman}
\node at (-1,0.45) {$g$};
\node at (-1,-0.45) {$g$};
\node at (1,0.55) {$O$};
\node at (1.0,-0.5) {$O$};
\end{tikzpicture}}\ \ +\ \ \scalebox{0.75}{\begin{tikzpicture}[baseline={([yshift=-0.9ex]current bounding box.center)},xshift=12cm]
\begin{feynman}[large]
\vertex (i1);
\vertex [right = 2cm of i1] (i2);
\vertex [above left=1.5 cm of i1] (v1);
\vertex [below left=1.5cm of i1] (v2);
\vertex [above right=1.5cm of i2] (v3);
\vertex [below right=1.5cm of i2] (v4);
\diagram* {
(i1) -- [ultra thick, gluon] (i2),
(v1) -- [ultra thick, gluon] (i1),
(v2) -- [ultra thick, gluon] (i1),
(i2) -- [ultra thick, scalar] (v3),
(i2) -- [ultra thick, scalar] (v4),
};
\end{feynman}
\node at (-1,0.45) {$g$};
\node at (-1,-0.45) {$g$};
\node at (1,0.4) {$g$};
\node at (3,0.55) {$O$};
\node at (3,-0.5) {$O$};
\end{tikzpicture}}\ \ +\ \ \scalebox{0.75}{\begin{tikzpicture}[baseline={([yshift=-0.9ex]current bounding box.center)},xshift=12cm]
\begin{feynman}[large]
\vertex (i1);
\vertex [below = 1.8cm of i1] (i2);
\vertex [left = 1.3cm of i1] (v1);
\vertex [right= 1.3cm of i1] (v2);
\vertex [left = 1.3cm of i2] (v3);
\vertex [right=1.3cm of i2] (v4);
\diagram*{
(i2) -- [ultra thick, scalar] (i1),
(v1) -- [ultra thick, gluon] (i1),
(i1) -- [ultra thick, scalar] (v2),
(v3) -- [ultra thick, gluon] (i2),
(i2) -- [ultra thick, scalar] (v4),
};
\end{feynman}
\node at (-1,-0.45) {$g^a_{\mu}$};
\node at (-1,-1.35) {$g^b_{\mu}$};
\node at (0.4,-0.9) {$O^e$};
\node at (1.,-0.4) {$O^c$};
\node at (1,-1.4) {$O^d$};
\end{tikzpicture}}\ \ +\ \ \scalebox{0.75}{\begin{tikzpicture}[baseline={([yshift=-0.9ex]current bounding box.center)},xshift=12cm]
\begin{feynman}[large]
\vertex (i1);
\vertex [below = 1.8cm of i1] (i2);
\vertex [left = 1.3cm of i1] (v1);
\vertex [right= 1.3cm of i1] (v2);
\vertex [left = 1.3cm of i2] (v3);
\vertex [right=1.3cm of i2] (v4);
\diagram*{
(i2) -- [ultra thick, scalar] (i1),
(v1) -- [ultra thick, gluon] (i1),
(i1) -- [ultra thick, scalar] (v2),
(v3) -- [ultra thick, gluon] (i2),
(i2) -- [ultra thick, scalar] (v4),
};
\end{feynman}
\node at (-1,-0.45) {$g^a_{\mu}$};
\node at (-1,-1.35) {$g^b_{\mu}$};
\node at (0.4,-0.9) {$O^e$};
\node at (1.,-0.4) {$O^d$};
\node at (1,-1.4) {$O^c$};
\end{tikzpicture}}\\[3.5ex]
&\text{(b)}\ \ \ \ \ \scalebox{0.75}{\begin{tikzpicture}[baseline={([yshift=-0.9ex]current bounding box.center)},xshift=12cm]
\begin{feynman}[large]
\vertex (i1);
\vertex [right = 2cm of i1] (i2);
\vertex [above left=1.5 cm of i1] (v1);
\vertex [below left=1.5cm of i1] (v2);
\vertex [above right=1.5cm of i2] (v3);
\vertex [below right=1.5cm of i2] (v4);
\diagram* {
(i1) -- [ultra thick, gluon] (i2),
(v1) -- [ultra thick, fermion] (i1),
(i1) -- [ultra thick, fermion] (v2),
(v2) -- [ultra thick] (i1),
(i2) -- [ultra thick, scalar] (v3),
(i2) -- [ultra thick, scalar] (v4),
};
\end{feynman}
\node at (-1,0.52) {$q$};
\node at (-1,-0.52) {$\bar{q}$};
\node at (1,0.4) {$g$};
\node at (3,0.55) {$O$};
\node at (3,-0.5) {$O$};
\end{tikzpicture}}
\end{align*}
\caption{Diagrams for scalar sgluon pair production due to (a) gluon fusion and (b) quark-antiquark annihilation. The diagrams for pseudoscalar pair production are given by replacing $O \to o$ everywhere.}
\end{figure}

Unlike for single sgluon production, both gluon fusion and quark-antiquark annihilation contribute significantly to pair production. The operators that open these production channels are given explicitly by \eqref{eA1}. The diagrams for these channels are provided in \hyperref[f4]{Figure 4}, with colors labeled where disambiguation is helpful.
The hadron-level cross sections of these production modes are given by
\begin{align}\label{e16}
\nonumber \sigma(gg \to OO) &= \int_{4m_O^2/s}^1 \d x_1 \int_{4m_O^2/sx_1}^1 \d x_2\, g(x_1,4m_O^2)g(x_2,4m_O^2)\, \hat{\sigma}(gg\to OO)\\
    \text{and}\ \ \ \sigma(\bar{q}q \to OO) &= \begin{multlined}[t][11.5cm]\int_{4m_O^2/s}^1 \d x_1 \int_{4m_O^2/sx_1}^1 \d x_2\, \bigg[\bar{f}(x_1,4m_O^2) f(x_2,4m_O^2)\\ + f(x_1,4m_O^2)\bar{f}(x_2,4m_O^2)\bigg]\, \hat{\sigma}(\bar{q}q \to OO),\end{multlined}
\end{align}
where $f(x,q^2)$ and $\bar{f}(x,q^2)$ are the quark and antiquark distribution functions, and where the parton-level cross sections are \cite{Choi:2009co}
\begin{align}\label{e17}
    \nonumber \hat{\sigma}(gg \to OO) &= \frac{15\pi}{16}\, \alpha_3^2\, \frac{1}{\hat{s}}\, \beta_O \left[1 +  \frac{34}{5}\frac{m_O^2}{\hat{s}} - \frac{24}{5}\left(1 - \frac{m_O^2}{\hat{s}}\right)\frac{m_O^2}{\hat{s}}\frac{1}{\beta_O} \ln \frac{1+\beta_O}{1-\beta_O}\right]\\
    \text{and}\ \ \ \hat{\sigma}(\bar{q}q \to OO) &= \frac{2\pi}{9}\, \alpha_3^2\, \frac{1}{\hat{s}}\, \beta_O^3
\end{align}
with $\hat{s} = x_1 x_2 s$. In these expressions, the kinematic function $\beta_O = [1- 4m_O^2 s^{-1}]^{1/2}$ is the speed of either sgluon in the pair, and we take the renormalization and factorization scales to be twice the sgluon mass. We emphasize that these expressions are written for the scalar but apply equally to the pseudoscalar. 
\section{Numerical results and phenomenology}
\label{s5}

In this section and the next we describe the phenomenology of the color-octet scalars in minimal $R$-symmetric models, using the analytic results presented in \hyperref[s4]{Section 4}, in the benchmark scenarios described in \hyperref[s3]{Section 3}. We first study the total decay rates, branching fractions, and characteristic traveling distances for the scalar and pseudoscalar sgluons and discuss the implications of these results for detection at the LHC. We then consider the production cross sections. In \hyperref[s6]{Section 6} we merge these two discussions to revisit constraints on these particles from searches conducted at the LHC.

\subsection{Decay rates and characteristic distances}\label{s5.A}

\begin{figure}\label{f5}
\begin{center}
\begin{tabular}{c c}
\text{(a)} & \includegraphics[align=c, trim=-2.4cm 0 0 0, width=.775\linewidth]{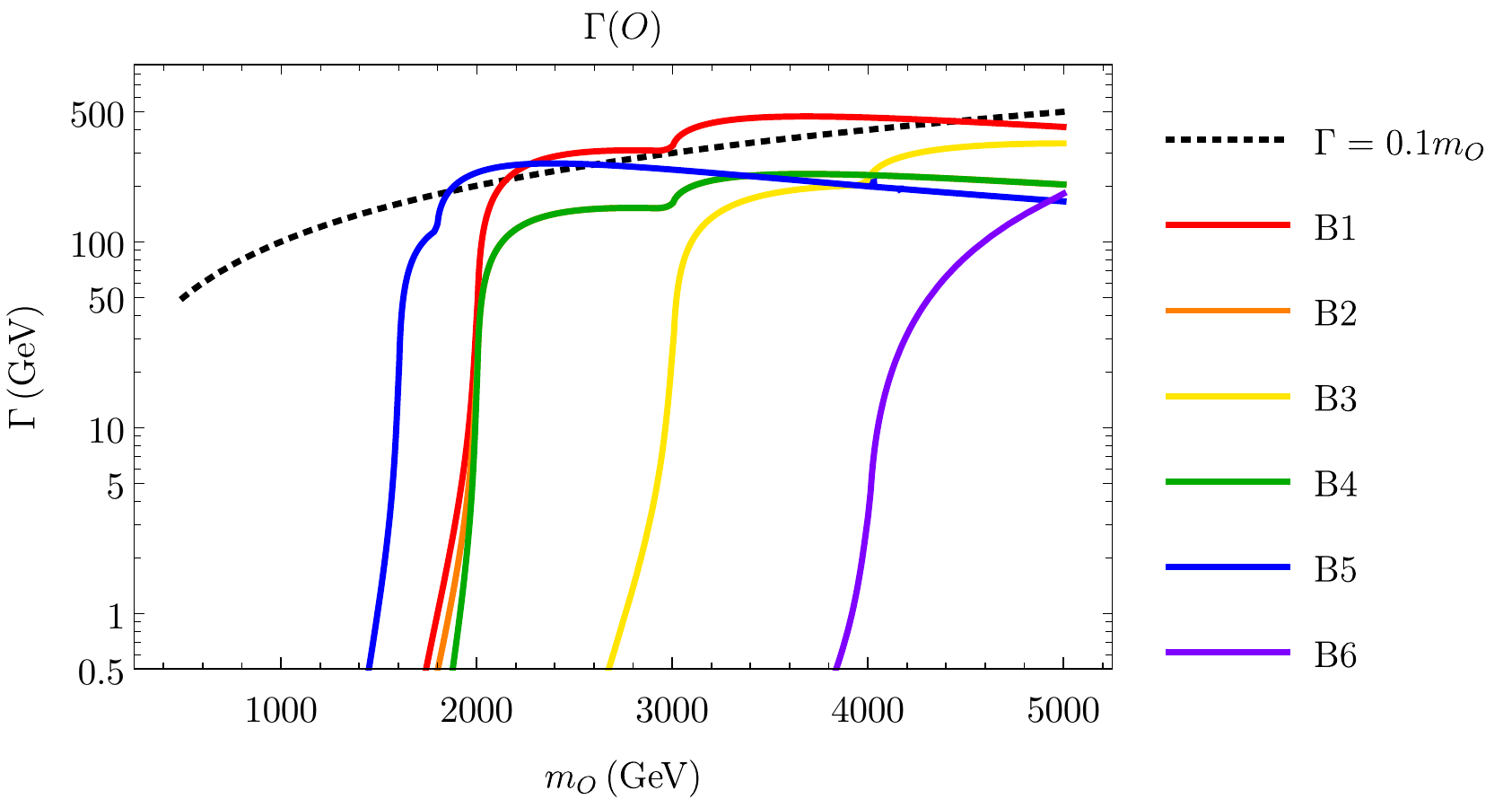}\\[5ex]
\rule{0pt}{20ex}\text{(b)} & \includegraphics[align=c, width=.6\linewidth]{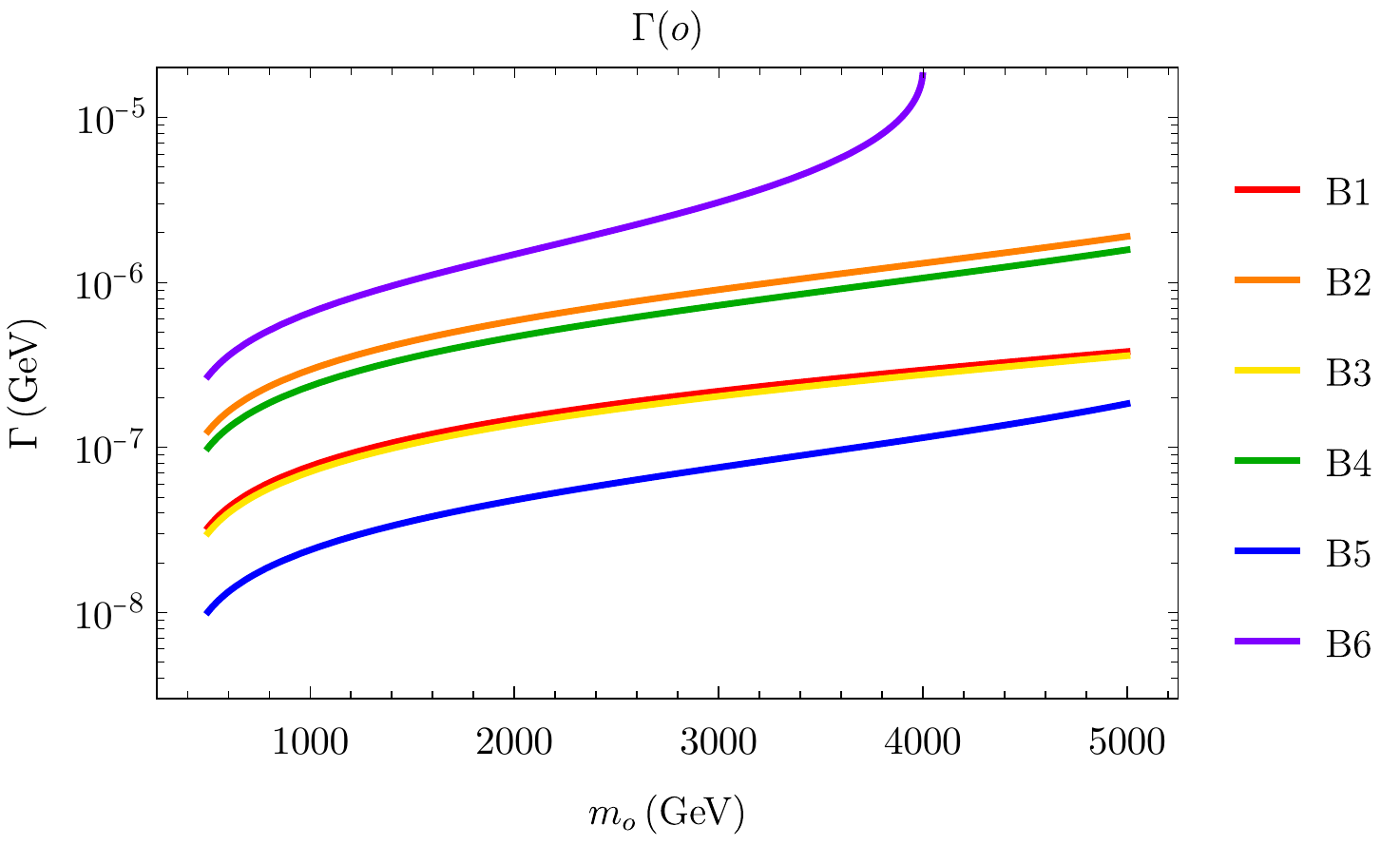}
\end{tabular}
\end{center}
\caption{Total decay widths in benchmarks B1--B6 of (a) the scalar sgluon and (b) the pseudoscalar. Scalar widths approach or exceed $10\%$ of the particle's mass above the squark decay threshold.}
\end{figure}

The total decay widths $\Gamma(O)$ and $\Gamma(o)$ of the sgluons are given by the appropriate sums of \eqref{e11}, \eqref{e12}, \eqref{eB}, \eqref{e13}, and \eqref{e14}. These are plotted in \hyperref[f5]{Figure 5}, in the six benchmarks displayed in \hyperref[tII]{Table 2}, as functions of the sgluons' masses. Numerical evaluation of the Passarino-Veltman functions \cite{Passarino:1979pv} was carried out here and subsequently using the \textsc{Mathematica}$^{\copyright}$\ package \textsc{Package-X} \cite{Mathematica, Patel:2017px}. Immediately we see a stark contrast between the sgluons: whereas the pseudoscalar has exceedingly small decay widths in all benchmarks across the entire mass range, the width of the scalar becomes quite large --- exceeding ten percent of the particle's mass --- once the particle can decay to on-shell squarks.

\begin{figure}\label{f6}
\begin{center}
\begin{tabular}{c c}
\text{(a)}\ \ \ \ \ \ \ \ \ \ \ & \includegraphics[align=c, width=.7\linewidth]{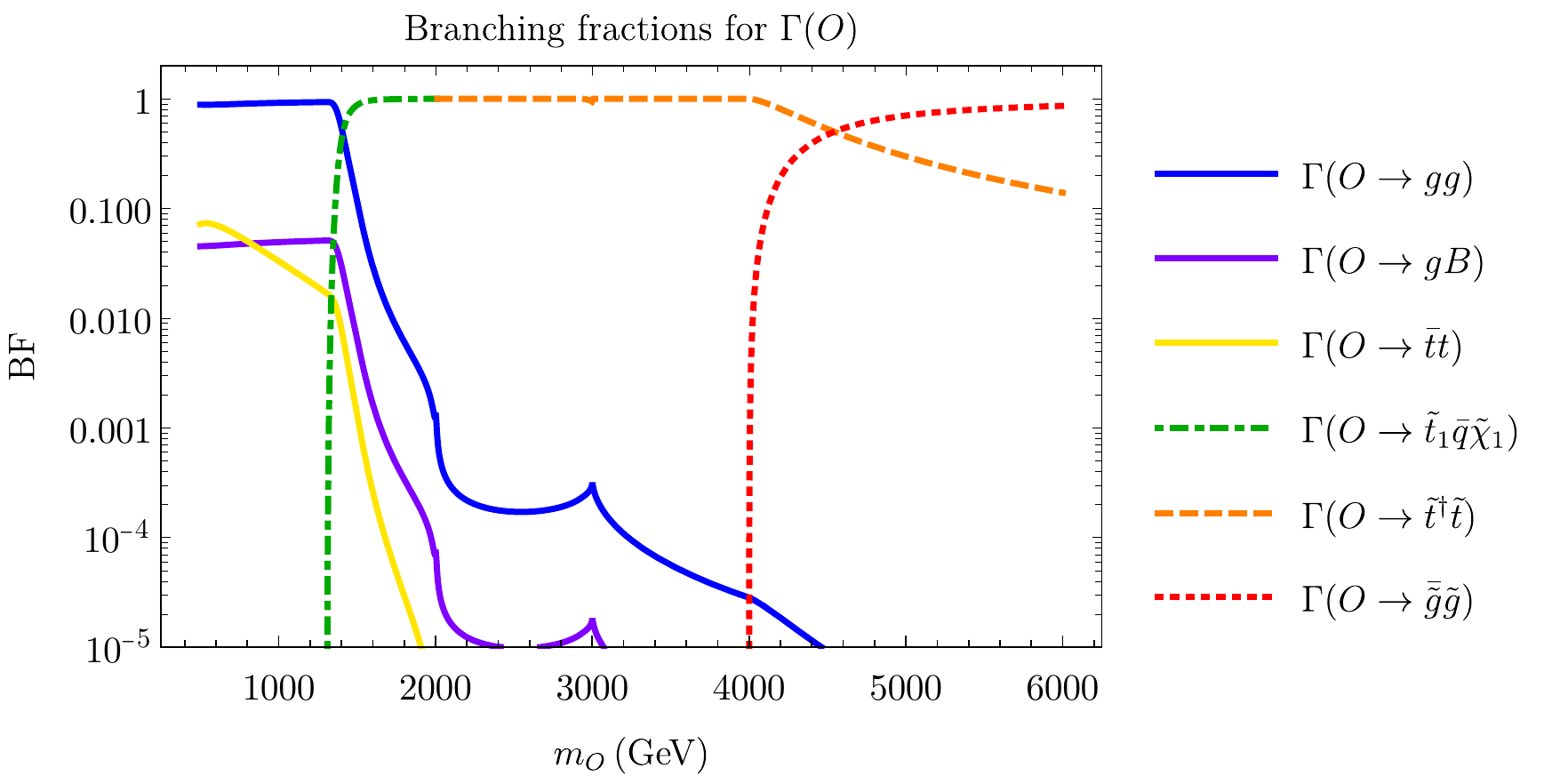}\\[5ex]
\rule{0pt}{20ex}\text{(b)}\ \ \ \ \ \ \ \ \ \ \ & \includegraphics[align=c, width=.7\linewidth]{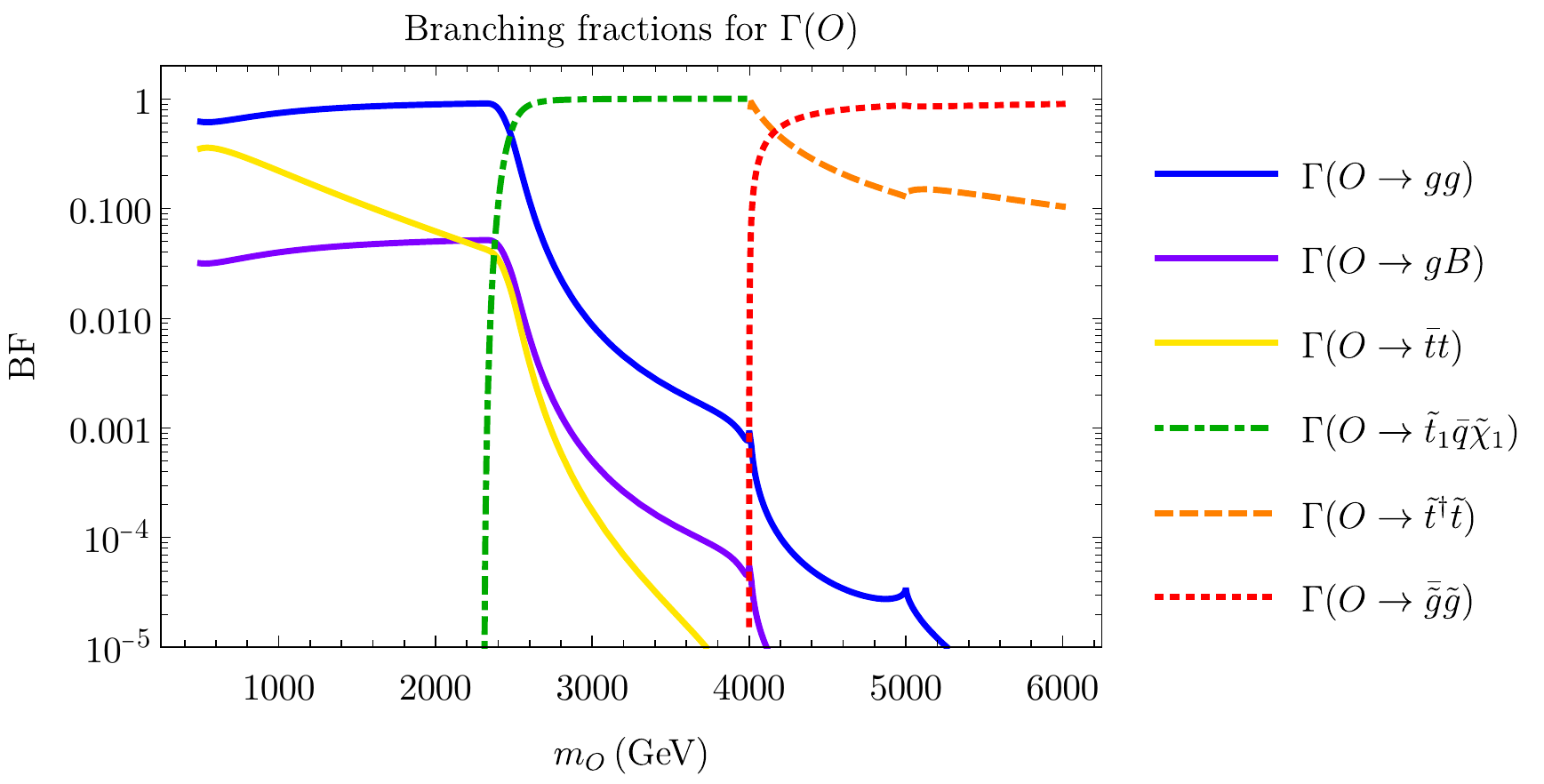}
\end{tabular}
\end{center}
\caption{Branching fractions for the scalar sgluon in (a) a modified B1 with $m_3 = 2.0\, \text{TeV}$ and (b) B6.}
\end{figure}

\begin{figure}\label{f7}
\begin{center}
\includegraphics[width=.48\textwidth]{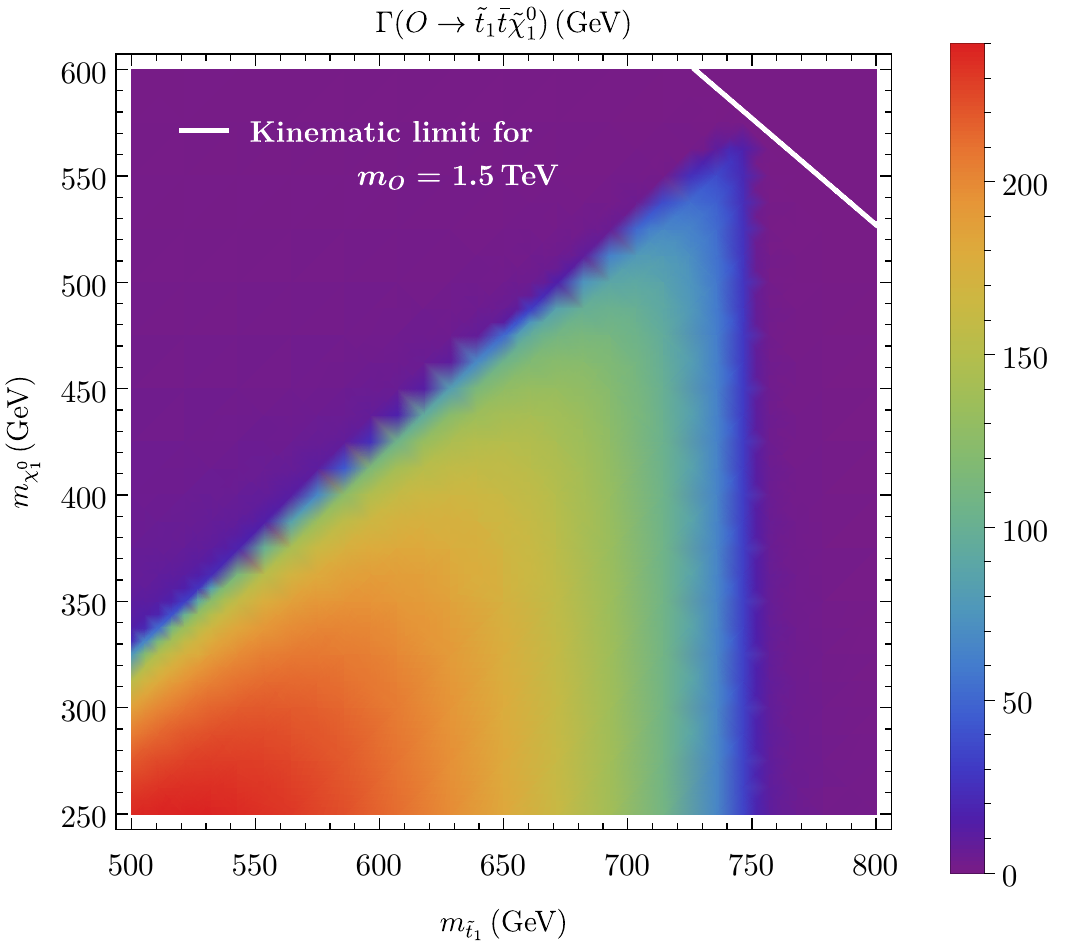}
\end{center}
\caption{Partial width of decay of a scalar sgluon of mass $m_O = 1.5\, \text{TeV}$ to a light stop, a top quark, and the Higgsino LSP.}
\end{figure}

The branching fractions for the decays of the scalar sgluon are plotted in \hyperref[f6]{Figure 6}. These plots combine the $g\gamma$ and $gZ$ channels into a channel denoted by $gB$ and combine the three-body decays beneath the on-shell squark decay threshold into a channel denoted by $\tilde{t}_1 \bar{q}\tilde{\chi}_1$. The plot in \hyperref[f6]{Figure 6(a)} is in a modified benchmark B1 in which the gluino mass is lowered to $2.0\, \text{TeV}$ to kinematically open that channel for sgluons lighter than $5.0\, \text{TeV}$. We do this to show that, once allowed, the gluino decay will come to dominate every other channel. The plot in \hyperref[f6]{Figure 6(b)} is in B6, where the gluino is degenerate with the light stop at $2.0\, \text{TeV}$. These plots have several features worth noting. First, we see that before the single squark mass threshold is reached, gluon decays dominate the branching fractions in B1. However, in the B6 benchmark with heavy stops, the decay to gluons is suppressed by the heavy stop mass such that the loop-induced decay to quarks achieves near parity with the decay to gluons. The mass of the sgluon is harder to constrain in scenarios where neither of the decay channels open to light sgluons can dominate. We elaborate upon this point in \hyperref[s6]{Section 6}. We next note that once kinematically allowed, the three-body decays dominate, beating the loop-level decays in all benchmarks. This decay channel is clearly supplanted by decays to squarks above the on-shell squark threshold.
We pause to note that both the threshold and the partial width of the three-body decay depend on the masses of the final-state particles. We demonstrate in \hyperref[f7]{Figure 7} the effects on \eqref{e14} of varying the light stop and LSP masses while holding the scalar mass fixed at $m_O = 1.5\, \text{TeV}$. Though we have not chosen such a benchmark here, we note that points in parameter space with fairly light stops are of some interest. For example, points like $(m_{\tilde{t}_1},m_{\tilde{\chi}_1^0}) = (500\, \text{GeV}, 300\, \text{GeV})$ --- which have not yet been ruled out for models with a Higgsino (N)LSP \cite{ATLAS:2018st} --- produce enhanced partial three-body decay widths at thresholds as low as $m_{\tilde{t}_1} + m_{\tilde{\chi}_1^0} + m_t \approx 973\, \text{GeV}$ (or, by extension, $m_{\tilde{t}_1} + m_{\tilde{\chi}_1^-} + m_b \approx 809\, \text{GeV}$). This fact will have consequences for future searches once more data allows the $m_O \geq 1.0\, \text{TeV}$ region to become accessible.

\begin{figure}\label{f8}
\begin{center}
  \includegraphics[width=.48\linewidth]{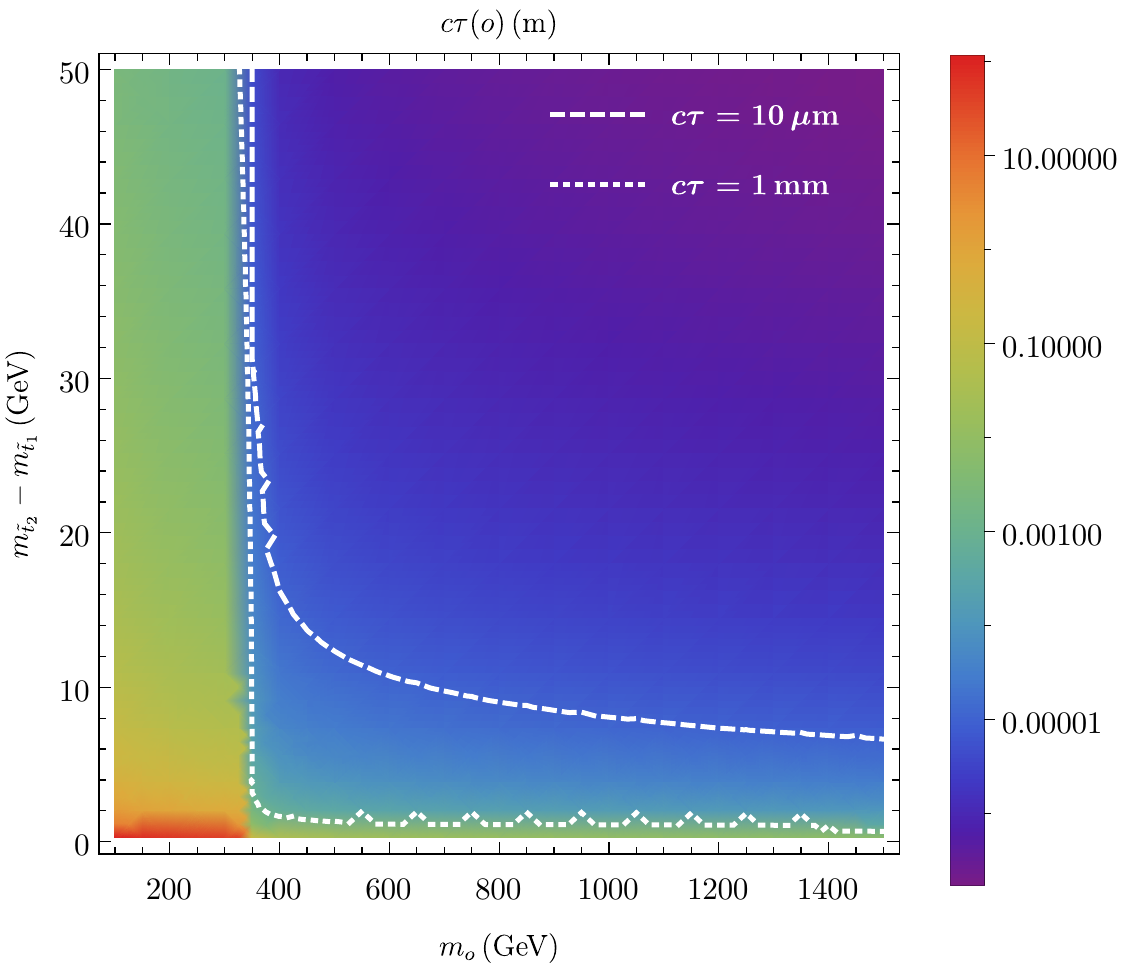}
\end{center}
\caption{Characteristic traveling distance of the pseudoscalar sgluon in a modified B1 with variable stop mass splitting.}
\end{figure}

It is instructive to compute the sgluons' characteristic traveling distances $c \tau(O)$ and $c\tau(o)$, with $\tau(X) = \hbar \Gamma^{-1}(X)$ the lifetime of a particle $X$. Since its width is orders of magnitude smaller than that of the scalar, the pseudoscalar generically travels farther in a detector than the scalar, but the discussion becomes interesting if the mass splitting between left- and right-chiral squarks is taken into account. Recall, in particular, that $\Gamma(O \to gg)$, $\Gamma(O \to g\gamma)$, $\Gamma(O \to gZ)$, and $\Gamma(O\, \text{and}\, o \to \bar{q}q)$ vanish if all squarks are degenerate. This cancellation --- which does not hold if the gluino is not Dirac (or a pair of degenerate Majorana fermions) and is slightly spoiled by electroweak contributions to the loop decays --- nevertheless suggests that there may be a region of parameter space, beneath the on-shell squark decay threshold, where at least the pseudoscalar sgluon may be long-lived. In order to explore this idea, we have plotted in \hyperref[f8]{Figure 8} the characteristic length of the pseudoscalar in a family of B1-like benchmarks with variable stop mass splitting.
Notice that a splitting of $\mathcal{O}(10)\, \text{GeV}$ is sufficient to allow even a moderately heavy pseudoscalar to travel a micron or more before decaying. We note that mass splittings around this size are of the same order as the Standard Model $D$ terms, which generically appear in the squark mass matrices. On the other hand, significantly smaller splittings --- of $\mathcal{O}(10^2)\, \text{MeV}$ --- are required in order to achieve long lifetimes for the scalar sgluon. Splitting of this size, quite small compared to the $\mathrm{TeV}$-scale stop masses considered in this work, seems unrealistic: even if degeneracy can be arranged at tree level, it is known that loop corrections to stop masses, while perhaps modest, should affect the stop mass splitting at the $\mathrm{GeV}$ level \cite{Donini:1996stoploop}. Therefore we do not take seriously the prospect of a long-lived \emph{scalar} sgluon.

Nevertheless, any model that predicts reasonably small mass splitting of left- and right-chiral states also predicts a \emph{pseudoscalar} long-lived enough to decay with displaced vertices \cite{CERN:2019ll}. This is an exciting signature that degrades the efficacy of most searches that constrain the pseudoscalar and calls for searches for long-lived particles (LLPs). We further recall from \hyperref[s4]{Section 4.1} that below the $\bar{t}t$ mass threshold, the pseudoscalar decay is totally dominated by the $\bar{b}b$ channel. In this region, the total decay rate is significantly smaller than the decay rate to tops due to suppression by the final-state quark mass (viz. \eqref{e13}). Thus we see in \hyperref[f8]{Figure 8} that under the $\bar{t}t$ mass threshold the pseudoscalar becomes generically long-lived for any squark mass splitting, and decays with extreme vertex displacements in the compressed squark mass region. We return to these ideas in \hyperref[s6]{Section 6}. 

\subsection{Production cross sections}\label{s5.B}

\begin{figure}\label{f9}
\begin{center}
\includegraphics[width=.6\linewidth]{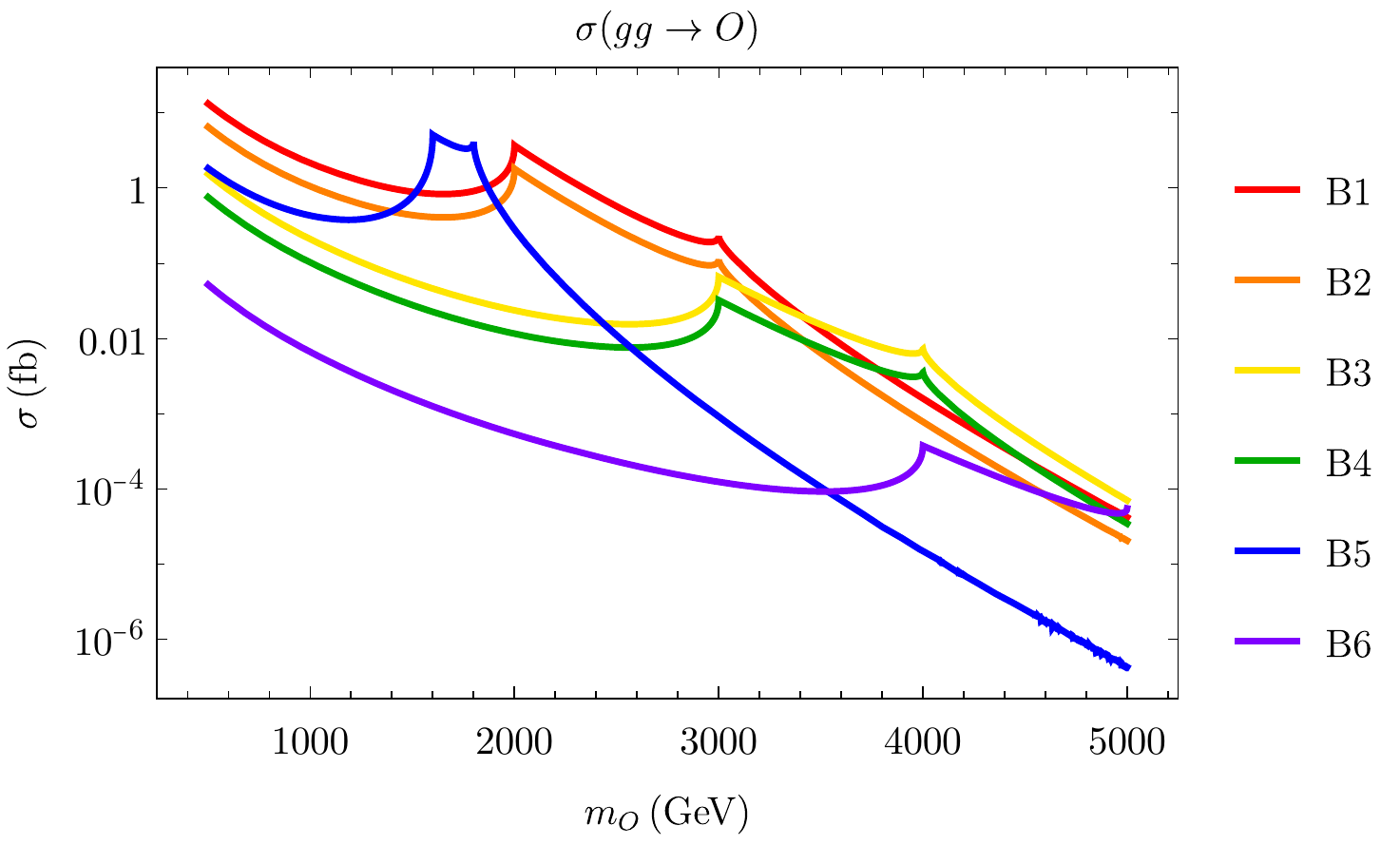}
\end{center}
\caption{Cross sections in benchmarks B1--B6 of single scalar sgluon production due to gluon fusion.}
\end{figure}

\begin{figure}\label{f10}
\begin{center}
\includegraphics[width=.725\linewidth]{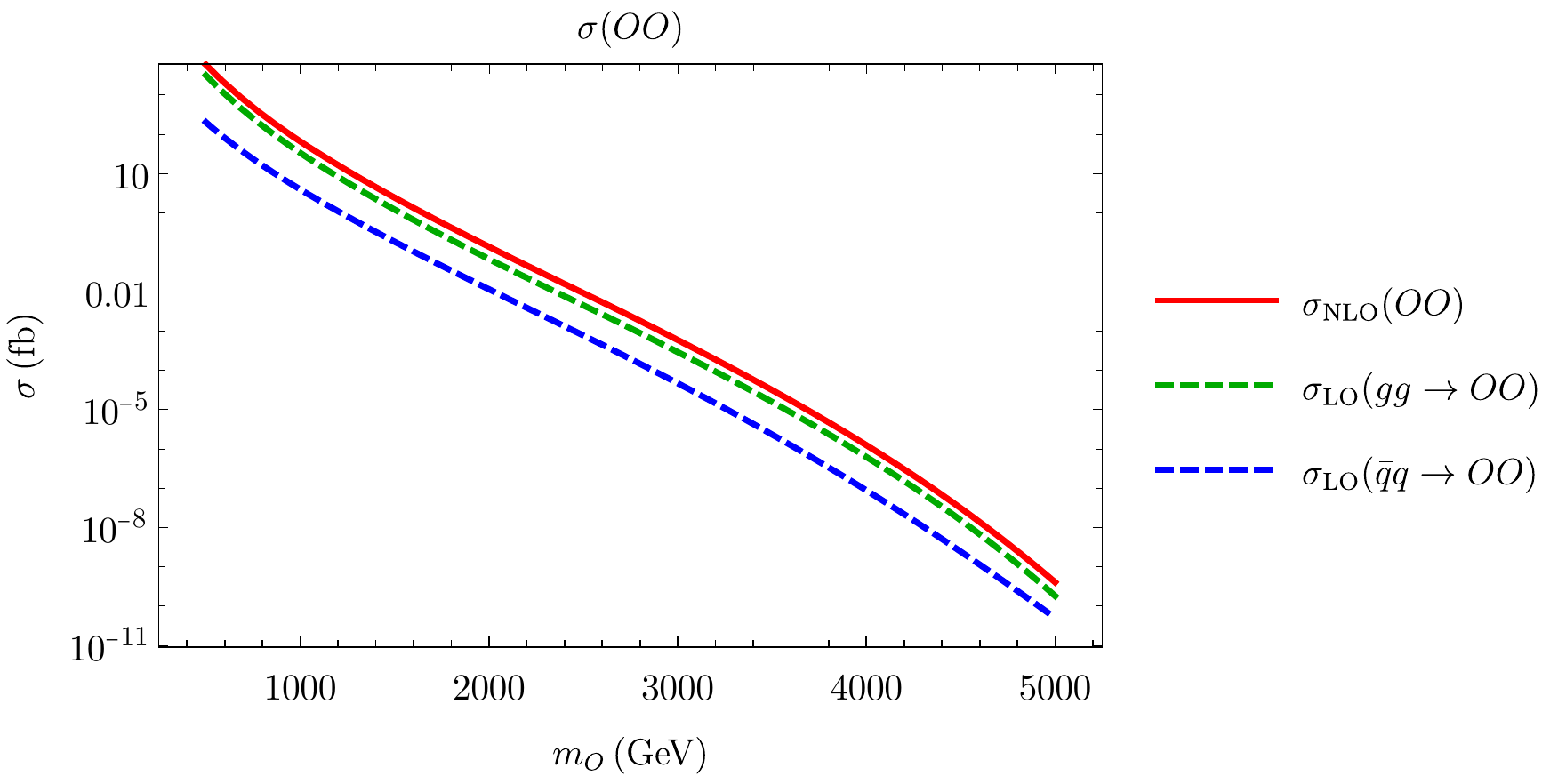}
\end{center}
\caption{Cross sections of either scalar or pseudoscalar sgluon pair production due to gluon fusion and quark-antiquark annihilation.}
\end{figure}

The cross section of single scalar sgluon production is given by \eqref{e15}. This cross section at $s^{1/2}=13\, \text{TeV}$ is plotted, in the six benchmarks displayed in \hyperref[tII]{Table 2}, in \hyperref[f9]{Figure 9}. This and subsequent plots, which required numerical integration of parton distribution functions, were generated using the \textsc{Mathematica}$^{\copyright}$\ package \textsc{ManeParse} to read the CT10 next-to-leading order (NLO) parton distribution functions \cite{Clark:2017mp, Lai:2010ct}.
The single sgluon production cross section has an interesting resonant enhancement that comes from crossing the on-shell mass threshold of the left- and right-chiral squark states. This can be seen most clearly in the B5 benchmark, whose enhancement region is narrow compared to the others' because the squarks in this scenario are split by only $100\, \text{GeV}$. The cross sections of scalar and pseudoscalar pair production are given by \eqref{e16}. These cross sections at $s^{1/2} = 13\, \text{TeV}$ are plotted in \hyperref[f10]{Figure 10}. Because the couplings that enable these production channels are determined by gauge invariance, the results are benchmark independent. Also plotted in this figure is an estimate of the total pair production cross section at NLO with a $K$-factor of $1.75$. This estimate is very close to the full NLO result for real color-octet scalars of mass $m_O = 1.0\, \text{TeV}$; it slightly overestimates the cross section beneath that mass, and we expect that it underestimates it above \cite{Degrande:2015pprod}. The total cross section attains a much higher peak than does the cross section of single sgluon production, but diminishes more precipitously. It is important to note that these results amount to half of the NLO cross section of pair production of \emph{complex} color-octet scalars \cite{Netto:2012nlo}, which are the targets of several LHC searches we discuss in \hyperref[s6]{Section 6}. We make clear in that section wherever we rescale theoretical cross sections to fit the models we study in this work.
\section{Revisiting collider constraints in realistic scenarios}\label{s6}

With the cross sections in hand, along with the sgluons' branching fractions, we can confront searches conducted at the LHC for color-octet scalars. As usual, because searches for beyond-Standard Model physics are most often interpreted for the MSSM or for simplified models, it is worth seeing how tightly our unique scenarios are constrained. In order to target the relevant searches, we recall from the previous section that only relatively light sgluons --- particles with $m_O\ \text{or}\ m_o \leq 1\, \mathrm{TeV}$ --- have appreciable production cross sections. In scenarios with light sgluons, the decays relevant to phenomenology are to two Standard Model gauge bosons and to third-generation quark-antiquark pairs.

\begin{figure}\label{f11}
\begin{center}
\includegraphics[width=.6\linewidth]{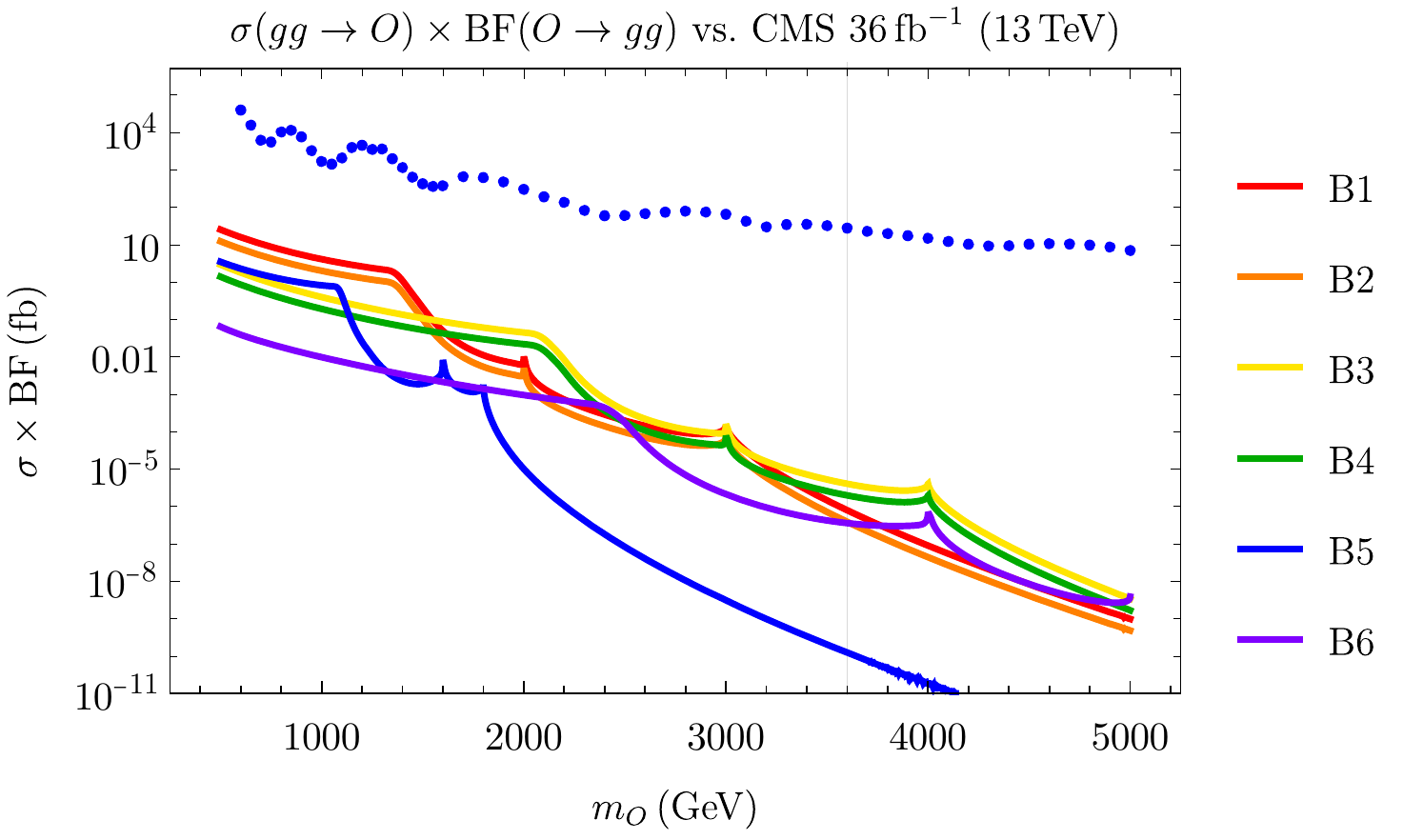}
\end{center}
\caption{Cross sections in benchmarks B1--B5 of single sgluon production with subsequent decay to gluons compared to exclusion bound from CMS $36\, \text{fb}^{-1}$ ($13\, \text{TeV}$) dijet resonance search for color-octet scalars assuming $\sigma(O) = \sigma_{\text{eff}}(gg \to O)$ \eqref{e18} and $\text{BF}(O \to gg)=1$. Exclusion is denoted by blue dots. Gray line marks CMS lower limit $m_O \geq 3.6\, \text{TeV}$.}
\end{figure}

We begin with a review of the most recent searches that have already been interpreted for models of color-octet scalars by the experimental collaborations, turning first to constraints on single scalar production. In \hyperref[f11]{Figure 11}, the cross sections for single scalar production in the six benchmarks discussed in \hyperref[s3]{Section 3} --- earlier plotted in \hyperref[f9]{Figure 9} --- are compared to the observed upper limits at $95\%$ confidence level (CL) \cite{Read:2002cls} from a CMS search at $s^{1/2} = 13\, \text{TeV}$ for dijet resonances \cite{CMS:2018s1}. CMS obtained these limits by interpreting their data for a benchmark color-octet scalar model assuming $\text{BF}(O \to gg) = 1$. This benchmark model explicitly assumes single color-octet scalar production via gluon fusion with a cross section given by
\begin{align}\label{e18}
    \sigma_{\text{eff}}(gg \to O) = \frac{5}{3} \pi^2\, \alpha_3\, k^2_{\text{eff}}\, \frac{1}{s} \int_{m_O^2/s}^1 \d x\, \frac{1}{x}\, g(x,m_O^2)g(m_O^2/sx,m_O^2)
\end{align}
with $k_{\text{eff}}^2 = 1/2$ \cite{Chivukula:2015ef}. This cross section significantly overestimates the cross sections of this production channel in our six benchmarks, even if the latter are enhanced at NLO (as in this figure) by a generous $K$-factor of $2.0$. Moreover, the branching fraction of the $gg$ decay in these models approaches unity only when the three-body decays are kinematically forbidden, and only then if the gluino far outweighs the stops. Accordingly, \hyperref[f11]{Figure 11} shows that scalar sgluons entirely evade the limits of this search in all six benchmarks.
Both because the total cross section is quite small, and since it is natural for $\text{BF}(O \to gg)$ to dominate $\text{BF}(O \to \bar{t}t)$, we expect cross sections of $\mathcal{O}(1)\, \text{fb}$ or smaller for boosted top quarks resulting from single scalar decay. Due to the low signal-to-background ratio, we do not expect constraints from searches for boosted top pairs to be significant. So altogether singly produced scalar sgluons are not tightly constrained.

\begin{figure}\label{f12} 
\begin{center}
\includegraphics[width=.6\linewidth]{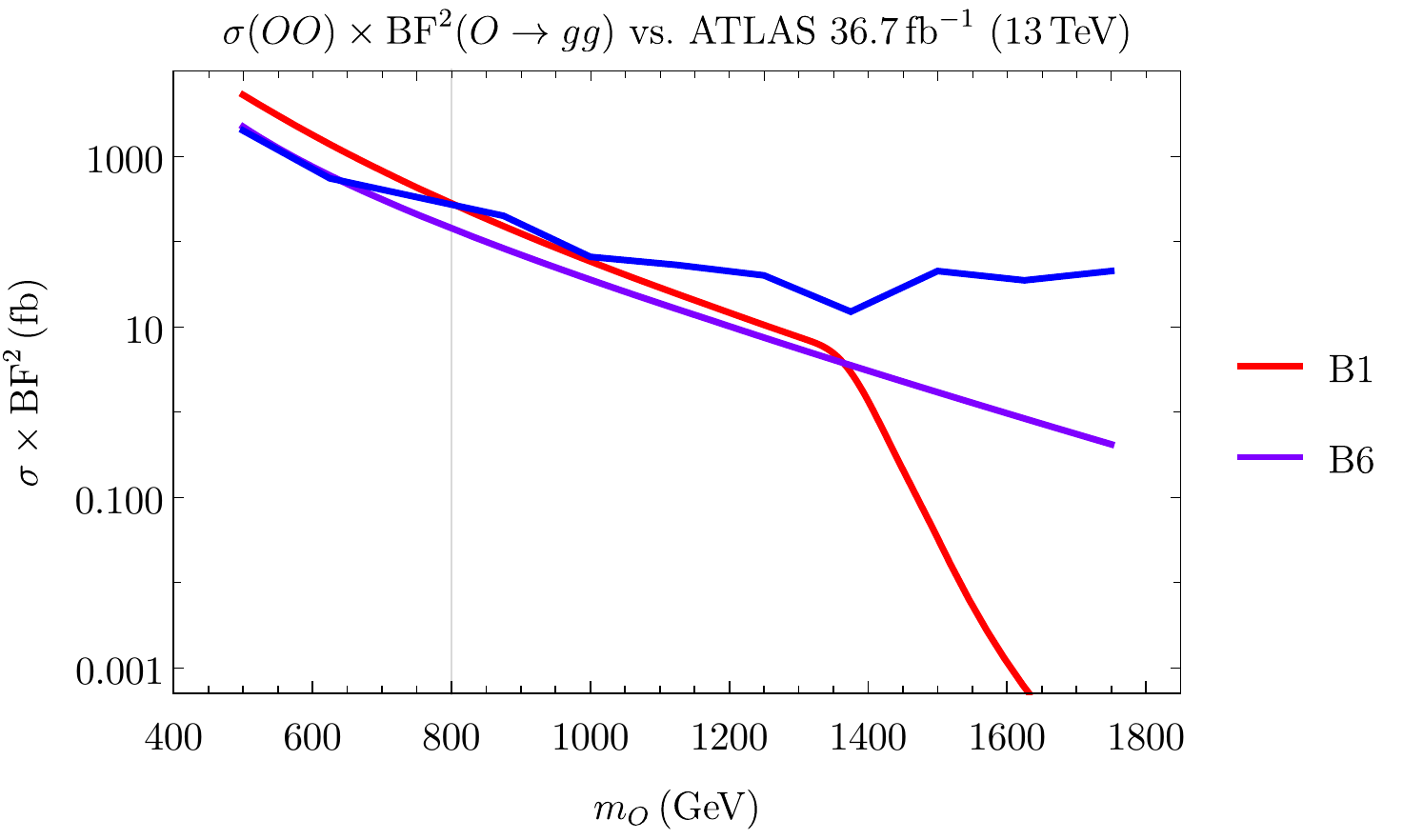}
\end{center}
\caption{Cross sections in benchmarks B1 and B6 of scalar sgluon pair production with subsequent decays to gluons compared to exclusion bound from ATLAS $36.7\, \text{fb}^{-1}$ ($13\, \text{TeV}$) search for color-octet scalars assuming $\text{BF}(O \to gg) = 1$. Exclusion is denoted by blue line. Gray line marks ATLAS lower limit $m_O \geq 800\, \text{GeV}$.}
\end{figure}

We now turn to constraints on sgluon pair production. In \hyperref[f12]{Figure 12}, the cross sections for scalar pair production --- earlier plotted in \hyperref[f10]{Figure 10} --- are compared to the observed upper limits at $95\%$ CL from an ATLAS search at $s^{1/2} = 13\, \text{TeV}$ for pair-produced resonances in flavorless four-jet final states \cite{ATLAS:2018s1}. ATLAS obtained these limits by interpreting their results for a model of real color-octet scalars assuming a pair production cross section very close to ours (\cite{Degrande:2015pprod}; viz. \hyperref[s5.B]{Section 5}) and $\text{BF}(O \to gg) = 1$. We note that the due to minimum $p_{\text{T}}$ cuts on jet energy, this search only applies to resonances with masses above $500\, \text{GeV}$. This search is unable to constrain the pseudoscalar sgluon, because this particle does not decay to gluons at one-loop level, but it can constrain the scalar in certain scenarios. We reinterpret this search for our color-octet scalars by accounting for different branching fractions. For instance, \hyperref[f12]{Figure 12} shows that the scalar in benchmark B1 is excluded for $m_O \in [500,800]\, \text{GeV}$ --- quite consistent with ATLAS' interpretation. On the other hand, in benchmark B6 --- where $m_{\tilde{t}_1} = m_3 = 2.0\, \text{TeV}$ and $m_{\tilde{t}_2} = 2.5\, \text{TeV}$ --- the upper end of the excluded region is lowered to $m_O \approx 640\, \text{GeV}$. The weakening of the exclusion is due to the fact that the $gg$ decay width drops with increasing squark mass as the branching fraction is taken up by the $\bar{q}q$ decay channels.

\begin{figure}\label{f13}
\begin{center}
\begin{tabular}{c c}
\text{(a)}\ \ \ \ \ \ \ \ \ \ \ & \includegraphics[align=c, width=.6\linewidth]{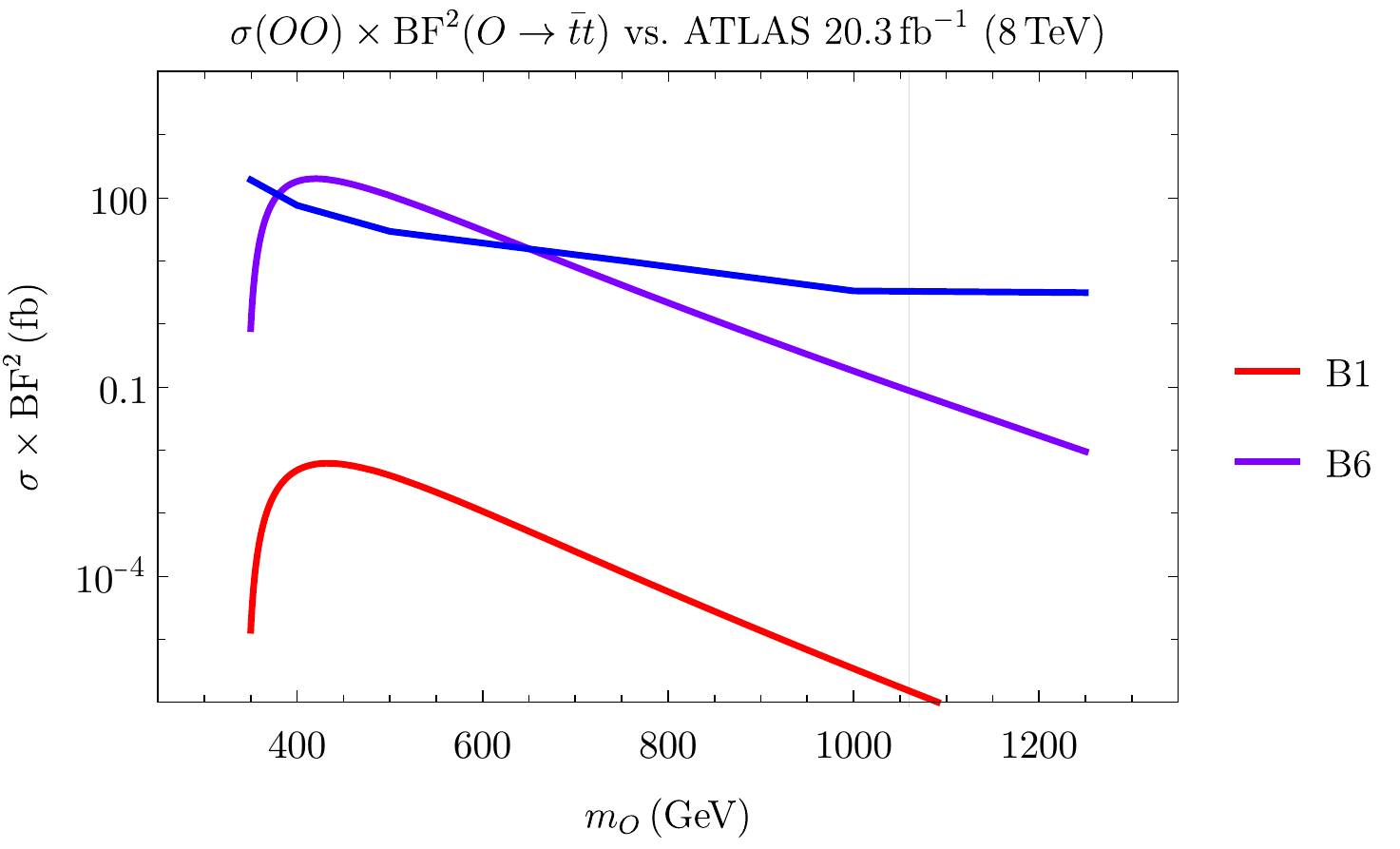}\\[5ex]
\rule{0pt}{20ex}\text{(b)}\ \ \ \ \ \ \ \ \ \ \ & \includegraphics[align=c, width=.6\linewidth]{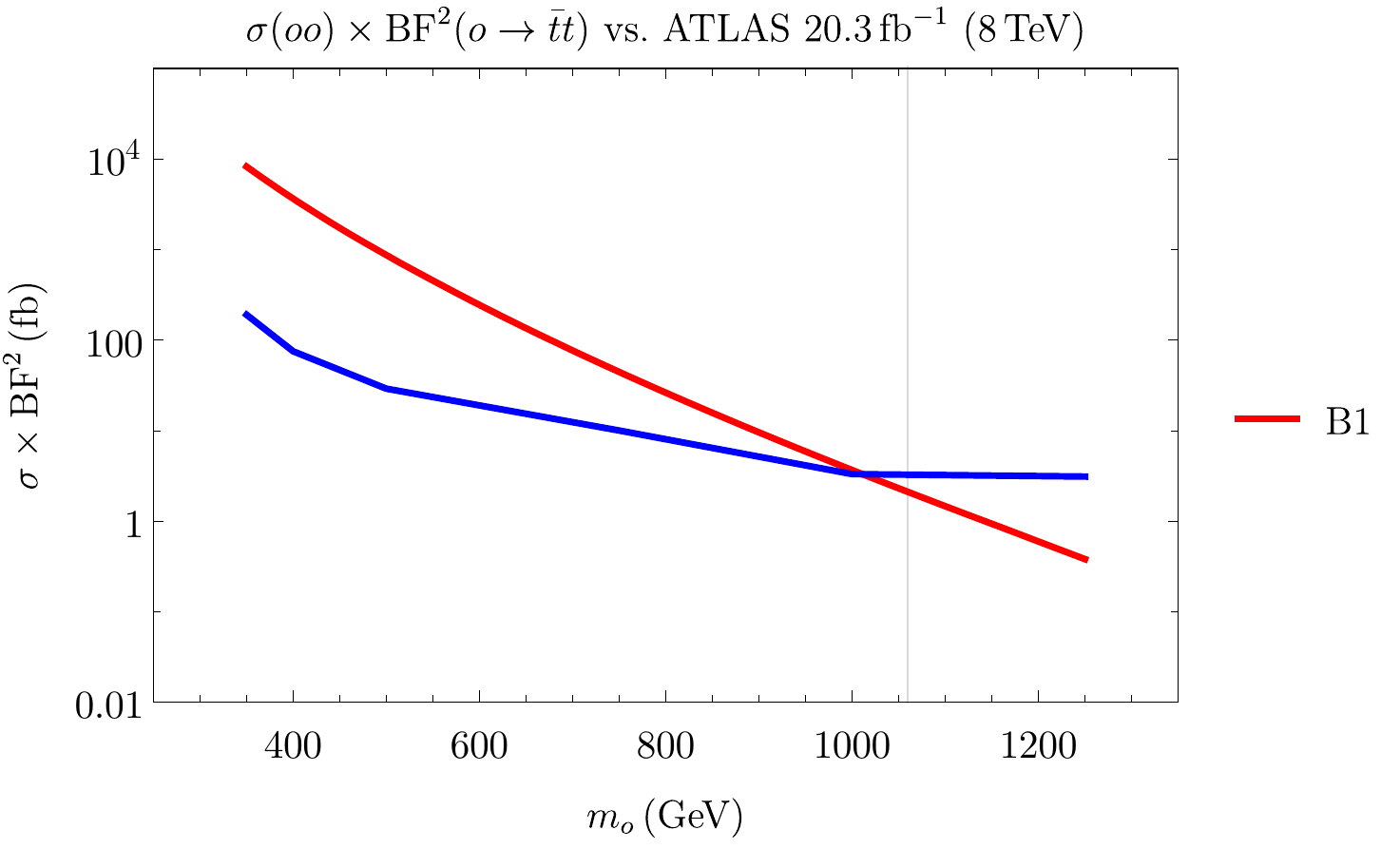}
\end{tabular}
\end{center}
\caption{Cross sections in benchmarks B1 and/or B6 of (a) scalar sgluon and (b) pseudoscalar pair production with subsequent decays to top-antitop pairs compared to exclusion bound from ATLAS $20.3\, \text{fb}^{-1}$ ($8\, \text{TeV}$) four-top event search for color-octet scalars assuming $\text{BF}(O\, \text{or}\, o \to \bar{t}t) = 1$. Exclusions are denoted by blue lines. Gray lines mark ATLAS lower limit $\sim\! 1.06\, \text{TeV}$ for the mass of a complex color-octet scalar.}
\end{figure}

In \hyperref[f13]{Figure 13}, we offer a similar comparison to the observed upper limits at $95\%$ CL from an ATLAS search at $s^{1/2} = 8\, \text{TeV}$ for four-top quark final states \cite{ATLAS:2015s2}. ATLAS obtained these limits by interpreting their results for a model of complex color-octet scalars assuming a cross section double ours (\cite{Netto:2012nlo}; viz. \hyperref[s5.B]{Section 5}) and $\text{BF}(O\, \text{or}\, o \to \bar{t}t) = 1$. This search only applies to resonances with masses above $350\, \text{GeV}$. We reinterpret this search for our real color-octet scalars by rescaling the cross section, assuming negligible differences in signal efficiencies, and by accounting for different branching fractions. This search can in principle constrain both scalar and pseudoscalar sgluons, since both decay at one-loop level to $\bar{t}t$. Like the other ATLAS search, however, this four-top search only has constraining power in part of parameter space. For example, the scalar sgluon in benchmark B1 easily evades the limits of this search, since in that scenario $\text{BF}(O \to gg) \gg \text{BF}(O \to \bar{t}t)$. In benchmark B6, on the other hand, the scalar is excluded for $m_O \in [400, 650]\, \text{GeV}$, as $\text{BF}(O \to \bar{t}t)$ is considerable in this scenario. This constraint naturally supersedes the more relaxed limit from the previous four-jet search. We explore the relationship between the two searches at the end of this section. Meanwhile, we note that a pseudoscalar with prompt decay is universally excluded by this search for $m_o \in [350,1011]\, \text{GeV}$, since this particle decays almost exclusively to $\bar{t}t$ beneath the threshold for decays to on-shell gluinos. Thus this search may bound pseudoscalars in regions of parameter space with sufficiently large splitting between left- and right-chiral squarks. We note that, due to its generically long lifetime in this region (viz. \hyperref[f8]{Figure 8}), pseudoscalars of mass $m_o < 350\, \text{GeV}$ will decay with large displaced vertices. As a result, searches for four-bottom quark final states that rely on prompt particle decay will not be sensitive. So four-top searches are the only relevant searches that assume prompt pseudoscalar decay.

We conclude with a more holistic look at the parameter space of $R$-symmetric models, including both older studies and searches that were not interpreted for color-octet scalars by the experimental collaborations, but have been recasted for these particles. As we do so, we account for the possible delayed decays of the pseudoscalar and extend our inquiry to sgluon masses beneath the ranges probed by the searches mentioned previously. Light scalar sgluons with $m_O \in [150, 270]\, \text{GeV}$ are excluded at 95\% CL by a somewhat older ATLAS search at $s^{1/2} = 7\, \text{TeV}$ for pair-produced colored resonances in four-jet final states \cite{ATLAS:2013s3}. These exclusion limits, which were obtained for a complex color-octet scalar model assuming a cross section double ours and $\text{BF}(O \to gg) = 1$, do not apply to the pseudoscalar but are robust when reinterpreted for the scalar, since the decay to $\bar{t}t$ is not accessible for $m_O \leq 350\, \text{GeV}$.

The exclusions we discussed above due to the $8\, \text{TeV}$ ATLAS four-top search have been rivaled \cite{Kotlarski:2016lep} or significantly improved by recasts of subsequent searches at $s^{1/2}=13\, \text{TeV}$ for jets and leptons \cite{ATLAS:2016lep} and production of four top quarks. We note in particular the recast \cite{Darme:2018rec} of a CMS measurement at $13\, \text{TeV}$ of the four-top quark ($\bar{t}t\bar{t}t)$ production cross section \cite{CMS:20184t}. The CMS measurement places an upper limit at $95\%$ CL of $\sim\! 30\, \text{fb}$ on beyond-Standard Model contributions to the $\bar{t}t\bar{t}t$ cross section, which (if interpreted as a conservative bound on the cross section of color-octet scalar pairs decaying to $\bar{t}t$) already supersedes the bounds of the $8\, \text{TeV}$ four-top search. This bound is strengthened for real color-octet scalars with mass between $800\, \text{GeV}$ and $1250\, \text{GeV}$ by the aforementioned recast. This measurement and its recast together extend the limit for sgluons decaying primarily to $\bar{t}t$ well into the TeV range.

\begin{figure}\label{f14}
\centering
\begin{subfigure}{.5\textwidth}
  \centering
\includegraphics[width=.95\linewidth]{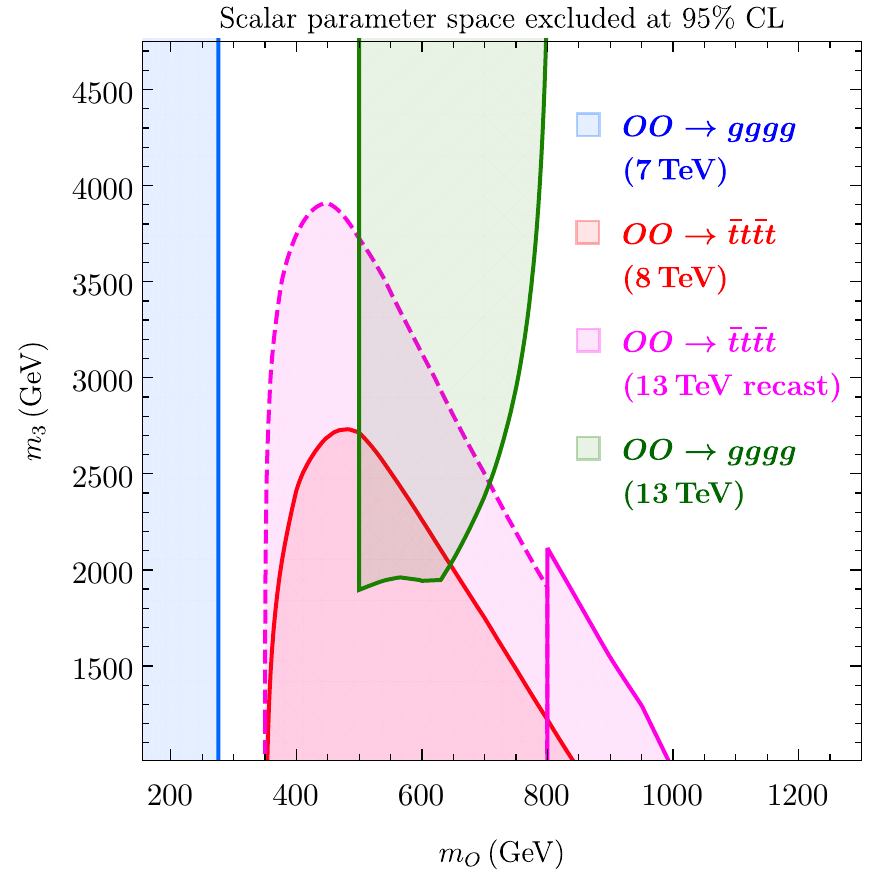}
\end{subfigure}%
\begin{subfigure}{.5\textwidth}
  \centering
  \includegraphics[width=.938\linewidth]{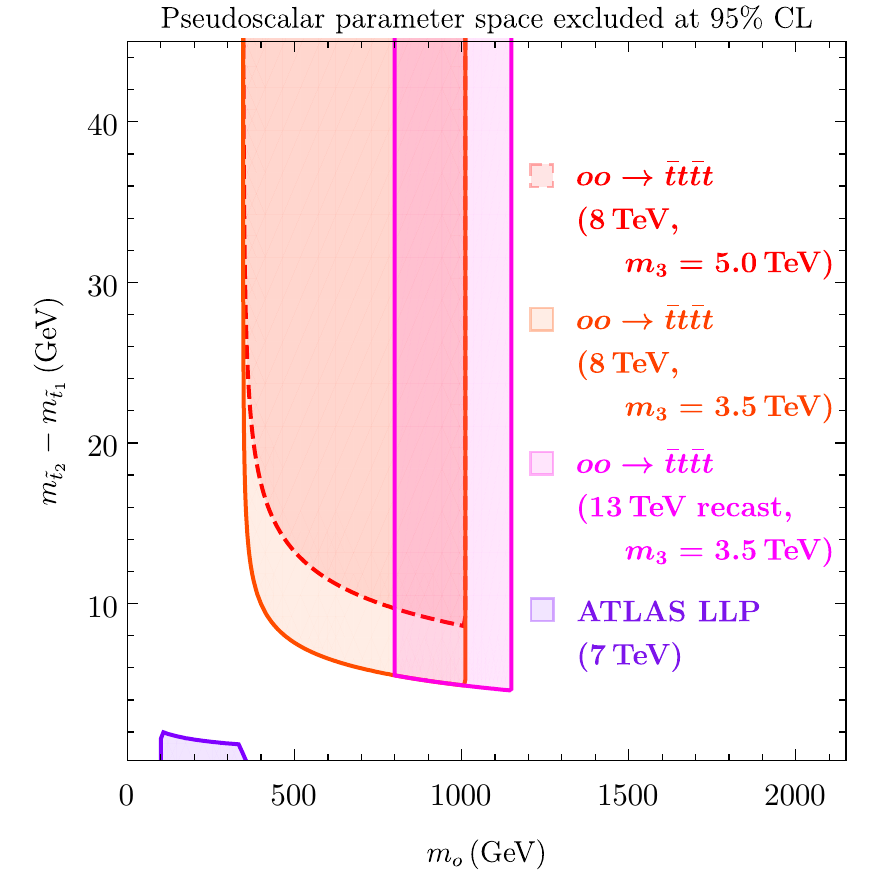}
\end{subfigure}
\caption{Excluded parameter space in (a, left) the $(m_O,m_3)$ plane for the scalar, with all other parameters in benchmark B6, and (b, right) the $(m_o,m_{\tilde{t}_2}-m_{\tilde{t}_1})$ plane for the pseudoscalar, with $m_{\tilde{t}_1}= 2.0\, \text{TeV}$ and $m_3 = 3.5\, \text{TeV}$ or $5.0\, \text{TeV}$.}
\end{figure}

The regions of scalar parameter space excluded by these searches, along with those excluded by the $8\, \text{TeV}$ four-top event search and $13\, \text{TeV}$ four-jet search, are plotted in \hyperref[f14]{Figure 14(a)}.
This plane of parameter space is in a family of B6-like benchmarks with variable gluino mass; here the lightest stop is fixed at $m_{\tilde{t}_1} =2.0\, \mathrm{TeV}$. Evident in this figure is a gap for at least $m_O \in (270, 350)\, \text{GeV}$ in which the scalar is unconstrained. Here the $\bar{t}t$ channel is not kinematically accessible, but the four-jet search does not have sensitivity. It may be quite difficult to close this gap by using similar four-jet searches at the LHC, given the low jet energy and multiplicity of these searches and the considerable chromodynamic background in this mass range. Also visible in this figure is the aforementioned interplay between the higher-energy pair production searches. In this mass range, where the only allowed decays are to Standard Model particles, the hierarchy between the gluino mass and the light squark mass controls which branching fraction ($\text{BF}(O \to gg)$ or $\text{BF}(O \to \bar{t}t)$) dominates, and therefore determines which searches (four-jet or four-top) constrain the scalar. For instance, when $m_3$ significantly exceeds $m_{\tilde{t}_2}$, $\Gamma(O \to gg) \gg \Gamma(O \to \bar{t}t)$ (viz. \hyperref[f6]{Figure 6(a)}) and the four-jet search provides the constraint. The joint exclusion limit from these searches can be relaxed to $m_O \approx 725\, \text{GeV}$ by simultaneously minimizing $\text{BF}(O \to gg)$ and $\text{BF}(O \to \bar{t}t)$.

The region of pseudoscalar parameter space excluded by the four-top event searches is plotted in \hyperref[f14]{Figure 14(b)}. This plane of parameter space is in a family of B1-like benchmarks with $m_{\tilde{t}_1} = 2.0\, \text{TeV}$ and with variable stop mass splitting. We take any point in this plane to be excluded by these searches only if the characteristic distance $c\tau(o)$ of the pseudoscalar is short enough for the decays of this particle to be considered prompt (viz. \hyperref[f8]{Figure 8}). This restriction reopens a considerable amount of parameter space, though we see that the excluded area expands again if the gluino mass is lowered. We also see that the recasted $13\, \text{TeV}$ four-top event search extends the limit on heavy promptly decaying pseudoscalars from $m_o \approx 1011\, \text{GeV}$ to $m_o \approx 1148\, \text{GeV}$.\footnote{This limit is a mild overestimate, since --- as we noted in \hyperref[s5.B]{Section 5} --- our choice of $K$-factor for NLO sgluon pair production mildly overestimates the cross section above $m_O$ or $m_o = 1\, \text{TeV}$.} These findings broadly suggest three scenarios: the pseudoscalar can be heavy and decay promptly; it can have moderate mass and be long-lived if the stops are nearly degenerate; finally, and most intriguingly, it can be very light with arbitrary splitting between squarks. A small amount of parameter space for $m_o \in [100, 350]\, \text{GeV}$ should be excluded in the latter scenario by an ATLAS search for long-lived hadronizing gluinos \cite{ATLAS:2011rh}. These exclusion limits, which are also included in \hyperref[f14]{Figure 14(b)}, strictly apply to gluinos long-lived enough to deposit energy in the tile calorimeter, but should apply to similarly long-lived pseudoscalar sgluons ($c\tau(o) \gtrsim 2.0\, \text{m}$) given the identical $\mathrm{SU}(3)_{\text{c}}$ representations and closely comparable pair-production cross sections \cite{Borschensky:2014gg}. Even accounting for searches for long-lived particles we clearly see that the pseudoscalar sgluon, in particular, is not tightly constrained.

Possibilities for closing gaps in parameter space may rely on more nonstandard search channels. For example, in the case of light scalar sgluons in the gap between $270\, \text{GeV}$ and $350\, \text{GeV}$, one strategy for the high-luminosity LHC (HL-LHC) run may be a new dedicated search exploiting the $g\gamma$ decay channel. The $g\gamma$ and $gZ$ channels have been shown to be powerful search tools for color-octet scalar models with single production modes \cite{Carpenter:2015gua}. Scalar pair production with one scalar decaying to $g\gamma$ may be a distinctive signature for detection, as it features a hard photon that would stand out above background.
\section{Outlook}
\label{s7}

In this work we have studied the color-octet scalars (sgluons) in minimal $R$-symmetric models. In addition to reviewing the known two-body decay modes of the scalar and pseudoscalar sgluons, we have extended the known result for the decay of a scalar to two gluons to cover decays to a gluon and a neutral electroweak gauge boson. Furthermore, we have presented an analytic expression for three-body decays involving a Higgsino (N)LSP whose branching fractions are non-negligible beneath the threshold for scalar decays to on-shell squarks. We have also reviewed the significant production modes of single scalars and pairs of either sgluon, and we have compared these results in a variety of interesting benchmarks to collider constraints on color-octet scalars. In so doing we have identified regions of parameter space where one or both sgluons can exhibit delayed or exotic decays, and we have demonstrated that exclusion limits from available searches for color-octet scalars at the Large Hadron Collider can be weakened or nullified for each sgluon in interesting and realistic scenarios. In short, the parameter space for the color-octet scalars in these models remains wide open.

It has been some time since the publication of the seminal studies of color-octet scalar collider phenomenology (\cite{Plehn:2008ae}, \cite{Choi:2009co}, and the like), and it is worth discussing for a moment how the models we have considered, and the results we have obtained, differ from the classic literature. Let us first review some important model differences. The analysis in \cite{Plehn:2008ae} is done in the MRSSM, which --- as we noted in \hyperref[s2]{Section 2} --- we set aside in favor of a simpler particle spectrum. In this sense, the minimal $R$-symmetric models we have studied hew more closely to the models explored in \cite{Choi:2009co}, which tacitly assume a level of $R$ symmetry sufficient to obtain Dirac gluinos but do not contain, e.g., $R$-Higgs bosons. But our models are not quite like those, either, since the latter assume an underlying $\mathcal{N}=2$ supersymmetry and we remain agnostic about this. We also assume that squark flavor mixing is negligible, which is consistent with \cite{Choi:2009co} but differs from \cite{Plehn:2008ae}. This restriction forbids mixed-flavor loop decays to $\bar{t}q$ or $\bar{q}t$, which were identified in \cite{Plehn:2008ae} as the decays (following sgluon pair production) with maximal discovery potential in the MRSSM. Finally, while the LSP is not discussed in \cite{Plehn:2008ae}, it is assumed in \cite{Choi:2009co} to be a photino. Both stand in contrast to our choice of Higgsino LSP, which we note can be naturally lighter than a bino or wino since the Dirac masses are naturally heavy. These differences, together with new benchmarks chosen to reflect more than a decade of exploration of supersymmetric parameter space, yield some different results and require somewhat different interpretation.

Nevertheless, some of the results we have presented are entirely consistent with the classic analyses, and a good deal of that work remains relevant today. The decays to quarks and gluons, including their peculiar behavior in the degenerate-squark limit, were thoroughly explored in these studies, as was the potentially copious single and pair production of sufficiently light color-octet scalars at the LHC. Moreover, the range of scalar sgluon masses identified in \cite{Plehn:2008ae} as ripe for exploration at the LHC is still of great interest today. In particular, as we discussed in \hyperref[s6]{Section 6}, the low end of that range ($m_O$ of $\mathcal{O}(10^2)\, \text{GeV}$) has been well probed except in the vicinity of the $\bar{t}t$ production threshold; meanwhile, as we have demonstrated, investigation of TeV-scale sgluons remains incomplete and worthwhile. With all of this in mind, we view the contribution of this work as threefold. First, we have clarified important phenomenological differences between the scalar and pseudoscalar sgluons that were not explored in \cite{Plehn:2008ae} (which drew few contrasts between these particles other than their mass difference) or \cite{Choi:2009co} (which took them to be degenerate and treated them as a complex scalar). Second, we have identified --- or more deeply explored --- decays unmentioned or only briefly noted in the older analyses, and we have discussed the possibility of exotic signatures in the event that the pseudoscalar sgluon is long-lived. Third, we have cataloged a number of experimental searches, performed in the years following the early theoretical studies, in order to show what parameter space (particularly what space explored in \cite{Plehn:2008ae}) remains open, and to see what scenarios should be investigated as we move forward.

We have done this work in anticipation of improved constraints on --- or perhaps discovery of --- color-octet scalars during future runs of the LHC, particularly after the HL-LHC upgrades currently scheduled for completion by the end of 2027. We naturally expect that enhanced luminosity will extend the collider's reach into regions of parameter space with higher sgluon masses. As we have discussed, it is possible that the scalar sgluon decay could be dominated by three-body decay channels in the $m_O \sim 1\, \text{TeV}$ region. Searches for these decays would be interesting, as they would feature multiple heavy-flavor quarks produced with significant missing transverse energy ($\slashed{E}{}_{\text{T}}$), as in $pp \to OO \to tt\bar{b}\bar{b}+\slashed{E}{}_{\text{T}}$ (here assuming that nearly degenerate charginos decay effectively invisibly to neutralinos). We also note that perhaps some extreme corners of parameter space may be explored in scenarios with more daring choices of LSP parameters (or NLSP in case of a gauge-mediated ultraviolet completion), which might markedly shift the lower bounds on squark masses. For example, a choice of sneutrino or stau NLSP may significantly alter the phenomenology. Other avenues of exploration include further study of long-lived hadronizing sgluons that present similarly to $R$-hadrons, albeit with opposite spin statistics. Depending on the details of their interactions, these may display a variety of novel signatures, and further long-lived particle searches may be able to constrain these scenarios.

\vfill

\pagebreak

\appendix

\section{Technical model details}
\label{aA}

In this appendix we discuss in detail the model framework, consistent with the discussion in Sections \hyperref[s2]{2} and \hyperref[s3]{3}, used to compute the analytic expressions in \hyperref[s4]{Section 4}. We discuss the strong sector and the electroweakinos separately.

\subsection*{Feynman rules for Dirac gluinos and color-octet scalars}

The Dirac gluino $\tilde{g}_3 \equiv \tilde{g}$ created by the coupling of the Majorana gluino $\lambda_3$ and the $\mathrm{SU}(3)_{\text{c}}$ adjoint Majorana fermion $\psi_3$ is written with explicit Weyl spinor indices $\{\alpha,\dot{\alpha}\}$ as
\begin{align}\label{eA0}
\tilde{g} = \begin{pmatrix}
\psi_{3\alpha}\\
\lambda_3^{\dagger \dot{\alpha}}
\end{pmatrix}\ \ \ \text{with charge conjugate}\ \ \ \tilde{g}^{\text{c}} \equiv \mathrm{C}\bar{\tilde{g}}_k^{\transpose} = \begin{pmatrix}
\lambda_{3\alpha}\\
\psi^{\dagger\dot{\alpha}}_3
\end{pmatrix},
\end{align}
where $\mathrm{C}$ is the charge-conjugation operator. The interactions between sgluons, quarks, gluons, squarks, and gluinos are governed by the following Lagrangian, derived from the supersoft operator \eqref{e1} and the K\"{a}hler potential \eqref{e5}:
\begin{multline}\label{eA1}
\mathcal{L}_O = \frac{1}{2}(\nabla_{\mu} O)^{\dagger}_a (\nabla^{\mu}O)^a + \frac{1}{2}(\nabla_{\mu} o)^{\dagger}_a (\nabla^{\mu}o)^a\\ - 2g_3 m_3O^a [\tilde{q}_{\text{L}}^{\dagger} \bt{t}_3^a \tilde{q}_{\text{L}} - \tilde{q}^{\dagger}_{\text{R}} \bt{t}_3^a \tilde{q}_{\text{R}}] - \ii g_3 f_{abc}\, O^a \bar{\tilde{g}}^b \tilde{g}^c + g_3f_{abc}\, o^a \bar{\tilde{g}}^b \gamma_5 \tilde{g}^c\\  - \sqrt{2}g_3\, \bar{q} \bt{t}^a_3 [\tilde{q}_{\text{L}} \text{P}_{\text{R}}\tilde{g}^a - \tilde{q}_{\text{R}} \text{P}_{\text{L}} \tilde{g}^{\text{c}a}]-\sqrt{2} g_3 [\tilde{q}_{\text{L}}^{\dagger}\, \bar{\tilde{g}}^a\text{P}_{\text{L}} - \tilde{q}^{\dagger}_{\text{R}}\, \overbar{\tilde{g}^{\text{c}}}^{\,a} \text{P}_{\text{R}}] \bt{t}^a_3 q,
\end{multline}
where $g_3$ is the $\mathrm{SU}(3)_{\text{c}}$ running coupling, $\bt{t}^a_3$ are the generators of the fundamental representation of $\mathrm{SU}(3)$, $f_{abc}$ are the structure constants of $\mathrm{SU}(3)$, and $\text{P}_{\text{L}}$ and $\text{P}_{\text{R}}$ are the left- and right-chiral projectors. The $\mathrm{SU}(3)_{\text{c}}$-covariant derivative $\nabla$ is defined by its action upon sgluons according to
\begin{align}
(\nabla^{\mu}O)^a = [\nabla^{\mu}]^a_{\ c} O^c = (\partial_{\mu}\delta^a_{\ c} + g_3 f^{ab}_{\ \ \,c}\, g^{\mu}_b)O^c,
\end{align}
where $g^{\mu}$ is a gluon field. The first line of \eqref{eA1} generates the interactions between sgluons and gluons that enable sgluon pair production. The corresponding Feynman rules are
\begin{align*}
\scalebox{0.75}{\begin{tikzpicture}[baseline={([yshift=-0.9ex]current bounding box.center)},xshift=12cm]
\begin{feynman}[large]
\vertex (i1);
\vertex [right = 1.5cm of i1] (i2);
\vertex [above left=1.5 cm of i1] (v1);
\vertex [below left=1.5cm of i1] (v2);
\diagram* {
(i1) -- [ultra thick, gluon] (i2),
(v2) -- [ultra thick, scalar,momentum'={[arrow shorten=0.25]$p_1$}] (i1),
(i1) -- [ultra thick, scalar, momentum'={[arrow shorten=0.25]$p_2$}] (v1),
};
\end{feynman}
\node at (-1.4,0.45) {$O^a\, \text{or}\, o^a$};
\node at (-1.4,-0.5) {$O^b\, \text{or}\, o^b$};
\node at (1.15,0.37) {$g^c_{\mu}$};
\end{tikzpicture}} = g_3 f_{abc} (p_1+p_2)_{\mu}\ \ \ \text{and}\ \ \ \scalebox{0.75}{\begin{tikzpicture}[baseline={([yshift=-.5ex]current bounding box.center)},xshift=12cm]
\begin{feynman}[large]
\vertex (i1);
\vertex [above left = 1.5cm of i1] (g1);
\vertex [below left = 1.5cm of i1] (g2);
\vertex [above right=1.5 cm of i1] (v1);
\vertex [below right=1.5cm of i1] (v2);
\diagram* {
(g2) -- [ultra thick, scalar] (i1),
(i1) -- [ultra thick, scalar] (g1),
(i1) -- [ultra thick, gluon] (v1),
(i1) -- [ultra thick, gluon] (v2),
};
\end{feynman}
\node at (-1.4,0.45) {$O^a\, \text{or}\, o^a$};
\node at (-1.4,-0.5) {$O^b\, \text{or}\, o^b$};
\node at (1.2,0.57) {$g^c_{\mu}$};
\node at (1.2,-0.525) {$g^d_{\nu}$};
\end{tikzpicture}} = \ii g_3^2 \eta_{\mu\nu}\, (f_{aec}f_{bed}+f_{bec}f_{aed}),
\end{align*}
where for simplicity we have not respected index height on the totally antisymmetric constants. The second and third lines of \eqref{eA1} generate interactions between sgluons, quarks, squarks, and gluinos, the Feynman rules for which are given below. Each field in each term in \eqref{eA1} is taken to flow into each vertex.
\begin{align*}
\scalebox{0.75}{\begin{tikzpicture}[baseline={([yshift=-.5ex]current bounding box.center)},xshift=12cm]
\begin{feynman}[large]
\vertex (i1);
\vertex [right = 1.5cm of i1] (i2);
\vertex [above right=1.5 cm of i2] (v1);
\vertex [below right=1.5cm of i2] (v2);
\diagram* {
(i1) -- [ultra thick, scalar] (i2),
(v2) -- [ultra thick, charged scalar] (i2),
(i2) -- [ultra thick, charged scalar] (v1),
};
\end{feynman}
\node at (2.75,0.75) {$\tilde{q}^{\dagger i}_{\text{L}}$};
\node at (2.75,-0.7) {$\tilde{q}_{\text{L}j}$};
\node at (0.3,0.3) {$O^a$};
\end{tikzpicture}} &= -2\ii g_3 m_3 [\bt{t}_3^a]_i^{\ j}\ \ \ \ \ \text{and}\ \ \ \ \ \scalebox{0.75}{\begin{tikzpicture}[baseline={([yshift=-.5ex]current bounding box.center)},xshift=12cm]
\begin{feynman}[large]
\vertex (i1);
\vertex [right = 1.5cm of i1] (i2);
\vertex [above right=1.5 cm of i2] (v1);
\vertex [below right=1.5cm of i2] (v2);
\diagram* {
(i1) -- [ultra thick, scalar] (i2),
(v2) -- [ultra thick, charged scalar] (i2),
(i2) -- [ultra thick, charged scalar] (v1),
};
\end{feynman}
\node at (2.75,0.75) {$\tilde{q}^{\dagger i}_{\text{R}}$};
\node at (2.75,-0.7) {$\tilde{q}_{\text{R}j}$};
\node at (0.3,0.3) {$O^a$};
\end{tikzpicture}} = 2\ii g_3 m_3 [\bt{t}_3^a]_i^{\ j},\\[5ex]
\scalebox{0.75}{\begin{tikzpicture}[baseline={([yshift=-.5ex]current bounding box.center)},xshift=12cm]
\begin{feynman}[large]
\vertex (i1);
\vertex [right = 1.5cm of i1] (i2);
\vertex [above right=1.5 cm of i2] (v1);
\vertex [below right=1.5cm of i2] (v2);
\diagram* {
(i1) -- [ultra thick, scalar] (i2),
(v2) -- [ultra thick, fermion] (i2),
(v2) -- [ultra thick, photon] (i2),
(i2) -- [ultra thick, fermion] (v1),
(i2) -- [ultra thick, photon] (v1),
};
\end{feynman}
\node at (2.75,0.7) {$\bar{\tilde{g}}^b$};
\node at (2.75,-0.7) {$\tilde{g}^c$};
\node at (0.3,0.3) {$O^a$};
\end{tikzpicture}} &= g_3 f_{abc}\ \ \ \ \ \ \ \ \ \ \ \ \ \ \ \text{and}\ \ \ \ \
\scalebox{0.75}{\begin{tikzpicture}[baseline={([yshift=-.5ex]current bounding box.center)},xshift=12cm]
\begin{feynman}[large]
\vertex (i1);
\vertex [right = 1.5cm of i1] (i2);
\vertex [above right=1.5 cm of i2] (v1);
\vertex [below right=1.5cm of i2] (v2);
\diagram* {
(i1) -- [ultra thick, scalar] (i2),
(v2) -- [ultra thick, fermion] (i2),
(v2) -- [ultra thick, photon] (i2),
(i2) -- [ultra thick, fermion] (v1),
(i2) -- [ultra thick, photon] (v1),
};
\end{feynman}
\node at (2.75,0.7) {$\bar{\tilde{g}}^b$};
\node at (2.75,-0.7) {$\tilde{g}^c$};
\node at (0.225,0.3) {$o^a$};
\end{tikzpicture}} = \ii g_3 f_{abc} \gamma_5,\\[5ex]
\scalebox{0.75}{\begin{tikzpicture}[baseline={([yshift=-.5ex]current bounding box.center)},xshift=12cm]
\begin{feynman}[large]
\vertex (i1);
\vertex [right = 1.5cm of i1] (i2);
\vertex [above right=1.5 cm of i2] (v1);
\vertex [below right=1.5cm of i2] (v2);
\diagram* {
(i1) -- [ultra thick, charged scalar] (i2),
(v2) -- [ultra thick, fermion] (i2),
(v2) -- [ultra thick, photon] (i2),
(i2) -- [ultra thick, fermion] (v1),
};
\end{feynman}
\node at (2.75,0.8) {$\bar{q}^i$};
\node at (2.75,-0.7) {$\tilde{g}^a$};
\node at (0.275,0.3) {$\tilde{q}_{\text{L}j}$};
\end{tikzpicture}} &= -\ii\sqrt{2}g_3 [\bt{t}_3^a]_i^{\ j} \text{P}_{\text{R}}\ \ \, \text{and}\ \ \ \scalebox{0.75}{\begin{tikzpicture}[baseline={([yshift=-.5ex]current bounding box.center)},xshift=12cm]
\begin{feynman}[large]
\vertex (i1);
\vertex [right = 1.475cm of i1] (i2);
\vertex [above right=1.5 cm of i2] (v1);
\vertex [below right=1.5cm of i2] (v2);
\diagram* {
(i1) -- [ultra thick, charged scalar] (i2),
(v2) -- [ultra thick, fermion] (i2),
(v2) -- [ultra thick, photon] (i2),
(i2) -- [ultra thick, fermion] (v1),
};
\end{feynman}
\node at (2.75,0.8) {$\bar{q}^i$};
\node at (2.775,-0.65) {$\tilde{g}^{\text{c}a}$};
\node at (0.25,0.3) {$\tilde{q}_{\text{R}j}$};
\end{tikzpicture}} = \ii\sqrt{2}g_3 [\bt{t}_3^a]_i^{\ j} \text{P}_{\text{L}},\\[5ex]
\text{and}\ \ \ \scalebox{0.75}{\begin{tikzpicture}[baseline={([yshift=-.5ex]current bounding box.center)},xshift=12cm]
\begin{feynman}[large]
\vertex (i1);
\vertex [right = 1.475cm of i1] (i2);
\vertex [above right=1.5 cm of i2] (v1);
\vertex [below right=1.5cm of i2] (v2);
\diagram* {
(i2) -- [ultra thick, charged scalar] (i1),
(i2) -- [ultra thick, fermion] (v2),
(v2) -- [ultra thick, photon] (i2),
(v1) -- [ultra thick, fermion] (i2),
};
\end{feynman}
\node at (2.75,0.75) {$q_j$};
\node at (2.75,-0.7) {$\bar{\tilde{g}}^a$};
\node at (0.34,0.35) {$\tilde{q}^{\dagger i}_{\text{L}}$};
\end{tikzpicture}} &= -\ii\sqrt{2}g_3 [\bt{t}_3^a]_i^{\ j} \text{P}_{\text{L}}\ \ \ \text{and}\ \ \ \scalebox{0.75}{\begin{tikzpicture}[baseline={([yshift=-.5ex]current bounding box.center)},xshift=12cm]
\begin{feynman}[large]
\vertex (i1);
\vertex [right = 1.5cm of i1] (i2);
\vertex [above right=1.5 cm of i2] (v1);
\vertex [below right=1.5cm of i2] (v2);
\diagram* {
(i2) -- [ultra thick, charged scalar] (i1),
(i2) -- [ultra thick, fermion] (v2),
(v2) -- [ultra thick, photon] (i2),
(v1) -- [ultra thick, fermion] (i2),
};
\end{feynman}
\node at (2.75,0.75) {$q_j$};
\node at (2.775,-0.6) {$\overbar{\tilde{g}^{\text{c}}}^{\,a}$};
\node at (0.34,0.35) {$\tilde{q}^{\dagger i}_{\text{R}}$};
\end{tikzpicture}} = \ii\sqrt{2}g_3 [\bt{t}_3^a]_i^{\ j} \text{P}_{\text{R}}.
\end{align*}

\subsection*{Feynman rules for Dirac Higgsinos}

Suppose for simplicity that the lightest neutralino is composed entirely of higgsinos $\tilde{H}_{\text{u}}^0$ and $\tilde{H}_{\text{d}}^0$, so that in the fashion of \eqref{eA0} we can write the lightest Dirac neutralino in terms of its Weyl components as
\begin{align}
\tilde{\chi}^0_1 = \begin{pmatrix}
\tilde{H}_{\text{u}\alpha}^0\\
\tilde{H}_{\text{d}}^{0 \dagger \dot\alpha}\end{pmatrix}.
\end{align}
In this case its mass originates at leading order from $\mathcal{L} \supset -\mu\, \overbar{\tilde{\chi}_1^0} \tilde{\chi}_1^0$, and we write the Feynman rules for the relevant interactions of the stops, a top quark, and the lightest neutralino as
\begin{align*}
\scalebox{0.75}{\begin{tikzpicture}[baseline={([yshift=-.5ex]current bounding box.center)},xshift=12cm]
\begin{feynman}[large]
\vertex (i1);
\vertex [right = 1.475cm of i1] (i2);
\vertex [above right=1.5 cm of i2] (v1);
\vertex [below right=1.5cm of i2] (v2);
\diagram* {
(i2) -- [ultra thick, charged scalar] (i1),
(i2) -- [ultra thick, fermion] (v2),
(v2) -- [ultra thick, photon] (i2),
(v1) -- [ultra thick, fermion] (i2),
};
\end{feynman}
\node at (2.75,0.8) {$t_i$};
\node at (2.75,-0.65) {$\overbar{\tilde{\chi}_1^0}$};
\node at (0.34,0.35) {$\tilde{t}^{\dagger i}_{1}$};
\end{tikzpicture}} &= -\ii y_t\, \text{P}_{\text{R}}\ \ \ \text{and}\ \ \ \scalebox{0.75}{\begin{tikzpicture}[baseline={([yshift=-.5ex]current bounding box.center)},xshift=12cm]
\begin{feynman}[large]
\vertex (i1);
\vertex [right = 1.475cm of i1] (i2);
\vertex [above right=1.5 cm of i2] (v1);
\vertex [below right=1.5cm of i2] (v2);
\diagram* {
(i2) -- [ultra thick, charged scalar] (i1),
(i2) -- [ultra thick, fermion] (v2),
(i2) -- [ultra thick, photon] (v2),
(v1) -- [ultra thick, fermion] (i2),
};
\end{feynman}
\node at (2.75,0.8) {$t_i$};
\node at (2.75,-0.65) {$\overbar{\tilde{\chi}_1^{0\text{c}}}$};
\node at (0.34,0.35) {$\tilde{t}^{\dagger i}_{2}$};
\end{tikzpicture}} = -\ii y_t \, \text{P}_{\text{L}},
\end{align*}
where $y_t$ is the top-quark Yukawa coupling. As mentioned in \hyperref[s3]{Section 3}, we take $\mu$ to be small enough (in accordance with its usual residence at the electroweak scale) that the extent of $R$ symmetry breaking induced by the $\mu$ term does not spoil the qualitative results that follow from assuming Dirac gauginos and unmixed stops. If the lightest chargino $\tilde{\chi}_1^{\pm}$ is also pure Higgsino, then at leading order it is degenerate with $\tilde{\chi}_1^0$. Furthermore, the Feynman rules for the interactions of the stops, a bottom quark, and the lightest chargino are the same as above with $t \to -b$ and $\tilde{\chi}_1^0 \to \tilde{\chi}_1^+$.
\section{Form factors for color-octet scalar decays}
\label{aB}

Here we provide some calculation details and explicit expressions for the form factors in the analytic partial decay rates in \hyperref[s4]{Section 4}. In addition to providing expressions (where possible) in terms of elementary functions, we express all loop integrals in terms of the scalar two- and three-point Passarino-Veltman functions \cite{Passarino:1979pv}
\begin{align*}
    B_0(p^2;M_1^2,M_2^2) &= \int \frac{\d^d \ell}{(2\pi)^d}\frac{1}{[\ell^2 - M_1^2][(\ell-p)^2 - M_2^2]}\\
    \text{and}\ \ \ C_0(p_1^2,(p_1+p_2)^2,p_2^2;M_1^2,M_2^2,M_3^2) &= \begin{multlined}[t][0cm]\\ \! \! \! \! \! \! \! \! \! \! \! \! \! \! \! \! \! \! \! \! \! \! \! \! \! \! \! \! \! \! \! \int \frac{\d^4 \ell}{(2\pi)^4} \frac{1}{[\ell^2-M_1^2][(\ell+p_1)^2-M_2^2][(\ell-p_2)^2-M_3^2]}.\end{multlined}
\end{align*}
Our $d$-dimensional integral measure $\d^d \ell\, (2\pi)^{-d}$ differs from the measure $\d^d \ell\, (\ii \pi^{d/2})^{-1}$ frequently used elsewhere, including in the original reference. In some places below, we exploit the symmetry of the three-point function under certain interchanges of its arguments, e.g. under $\{p_1^2 \leftrightarrow (p_1+p_2)^2, M_1^2 \leftrightarrow M_3^2\}$.

\subsection*{$\mathcal{F}(O \to gg)$: scalar decay to gluons}

The amplitude for this decay, assuming no squark mixing, can be written as
\begin{align*}
\mathcal{M}(O \to gg) = -\frac{g_3^3}{(4\pi)^2}\frac{m_3}{m_O^2}\, \varepsilon^*_{\nu}(k_1) \varepsilon^*_{\mu}(k_2)\, d_{abc} \left[m_O^2\eta^{\mu\nu} - 2 k_1^{\mu}k_2^{\nu}\right] \mathcal{F}(O \to gg),
\end{align*}
where the momenta and Lorentz indices are consistent with \hyperref[f2]{Figure 2(a)}, and where $\varepsilon^*_{\nu}(k_1)$ and $\varepsilon^*_{\mu}(k_2)$ are the gluon polarization vectors. The form factor $\mathcal{F}(O \to gg)$ is given in terms of Passarino-Veltman functions by
\begin{align}\label{eB1}
\mathcal{F}(O \to gg) = \begin{multlined}[t][10cm]32\ii \pi^2 \sum_{\tilde{q}} \bigg[m_{\tilde{q}_{\text{L}}}^2 C_0(m_O^2,0,0; m_{\tilde{q}_{\text{L}}}^2,m_{\tilde{q}_{\text{L}}}^2,m_{\tilde{q}_{\text{L}}}^2)\\ - m_{\tilde{q}_{\text{R}}}^2 C_0(m_O^2,0,0; m_{\tilde{q}_{\text{R}}}^2,m_{\tilde{q}_{\text{R}}}^2,m_{\tilde{q}_{\text{R}}}^2)\bigg],\end{multlined}
\end{align}
where the sum is over squark flavors. There is a sign difference in the form factor because left- and right-handed squarks couple to scalar sgluons with opposite charge; this is the origin of the cancellation of $\Gamma(O \to gg)$ for degenerate squarks. This form factor has a convenient representation in terms of elementary functions:
\begin{align}\label{eB2}
\mathcal{F}(O \to gg) = \sum_{\tilde{q}} \left[\tau_{\text{L}}f(\tau_{\text{L}}) - \tau_{\text{R}}f(\tau_{\text{R}})\right]_{\tilde{q}}\ \ \ \text{with}\ \ \ \tau_{\text{L}/\text{R}} = 4 \left(\frac{m_{\tilde{q}_{\text{L}/\text{R}}}}{m_O}\right)^2,
\end{align}
where
\renewcommand\arraystretch{1}
\begin{align}\label{eB3}
f(\tau)= \begin{cases}
-\dfrac{1}{4} \left[\ln \dfrac{1+\sqrt{1-\tau}}{1-\sqrt{1-\tau}} - \ii \pi\right]^2, & \tau < 1,\\[12.5pt]
\left[\sin^{-1} \dfrac{1}{\sqrt{\tau}}\right]^2, & \tau \geq 1.\end{cases}
\end{align}
In the heavy-squark limit, we have $\tau \gg 1$ and
\begin{align*}
\tau f(\tau) = \tau \left[\tau^{-1/2} + \frac{1}{6}\tau^{-3/2} + \mathcal{O}(\tau^{-5/2})\right]^2 = 1 + \mathcal{O}(\tau^{-1}),
\end{align*}
and the decay rate reduces to some multiple of the prefactor in \eqref{e12}, still vanishing if the squarks are degenerate.

\subsection*{$\mathcal{F}(O \to \bar{q}q)$: scalar decay to quarks}

The amplitudes for the part of this decay represented by the first and third diagrams in \hyperref[f2]{Figure 2(b)}, assuming no squark mixing, can be written for a single squark flavor in terms of Feynman parameter integrals as
\begin{multline*}
\mathcal{M}^{(13)}(O \to \bar{q}q) = \frac{2}{3} \frac{g_3^3}{(4\pi)^2}\, m_3m_q\, \bt{t}^a\\ \times \int_{\text{F}} \bar{u}(p_1,\sigma_1) \left[\frac{1}{\Delta_1(y,z)}\, (y \text{P}_{\text{L}} + z\text{P}_{\text{R}})-\frac{1}{\Delta_3(y,z)}\, (y\text{P}_{\text{R}} + z\text{P}_{\text{L}})\right]v(p_2,\sigma_2),
\end{multline*}
where $\bar{u}(p_1,\sigma_1)$ and $v(p_2,\sigma_2)$ are external fermion spinors, and where
\begin{align*}
\int_{\text{F}} = \int_0^1 \d z \int_0^{1-z}\d y
\end{align*}
denotes the integral over the functions
\begin{align}\label{eB4}
\nonumber \Delta_1(y,z) &= (1-y-z)m_3^2 + (y+z)m_{\tilde{q}_{\text{L}}}^2 - y z m_O^2 -(y+z)(1-y-z)m_q^2\\
\text{and}\ \ \ \Delta_3(y,z) &= (1-y-z)m_3^2 + (y+z)m_{\tilde{q}_{\text{R}}}^2 - y z m_O^2 -(y+z)(1-y-z)m_q^2
\end{align}
with $m_q$ the masses of the external quarks. The amplitudes for the second and fourth diagrams in that figure can similarly be written as
\begin{multline*}
\mathcal{M}^{(24)}(O \to \bar{q}q) = 3\, \frac{g_3^3}{(4\pi)^2}\, m_3 m_q\, \bt{t}^a\\ \times \int_{\text{F}} \bar{u}(p_1,\sigma_1)\, \bigg\lbrace\frac{1}{\Delta_2(y,z)}\, [(1-2y)\text{P}_{\text{L}} + (1-2z)\text{P}_{\text{R}}]\\ - \frac{1}{\Delta_4(y,z)}\, [(1-2y)\text{P}_{\text{R}} + (1-2z)\text{P}_{\text{L}}]\bigg\rbrace\, v(p_2,\sigma_2),
\end{multline*}
where now the integral is over the functions
\begin{align}\label{eB5}
\nonumber \Delta_2(y,z) &= (y+z)m_3^2 + (1-y-z)m_{\tilde{q}_{\text{L}}}^2 - y z m_O^2 - (y+z)(1-y-z)m_q^2\\
\text{and}\ \ \ \Delta_4(y,z) &= (y+z)m_3^2 + (1-y-z)m_{\tilde{q}_{\text{R}}}^2 - y z m_O^2 - (y+z)(1-y-z)m_q^2.
\end{align}
The form factor $\mathcal{F}(O \to \bar{q}q)$ can be written accordingly as
\begin{multline}\label{eB6}
\mathcal{F}(O \to \bar{q}q) = \sum_{\tilde{q}}\int_0^1 \d z \int_0^{1-z} \d y\\ \left\lbrace\frac{1}{9}(y+z)\left[\frac{1}{\Delta_1(y,z)}-\frac{1}{\Delta_3(y,z)}\right] +(1-y-z)\left[\frac{1}{\Delta_2(y,z)}-\frac{1}{\Delta_4(y,z)}\right]\right\rbrace.
\end{multline}
This form factor is also given in terms of Passarino-Veltman functions by
\begin{multline}\label{eB7}
    \mathcal{F}(O \to \bar{q}q) = 16\ii \pi^2 \sum_{q} \frac{1}{m_O^2-4m_q^2}\\ \times \left\lbrace \frac{1}{9} \left[\mathcal{I}^{(1)}_{\text{L}}(O \to \bar{q}q) -\mathcal{I}^{(1)}_{\text{R}}(O \to \bar{q}q)\right] + \left[\mathcal{I}^{(2)}_{\text{L}}(O \to \bar{q}q)  - \mathcal{I}_{\text{R}}^{(2)}(O \to \bar{q}q)\right]\right\rbrace
\end{multline}
with
\begin{align}\label{eB8}
\nonumber    \mathcal{I}_{\text{L}/\text{R}}^{(1)}(O \to \bar{q}q) &= \begin{multlined}[t][10cm] 2B_0(m_O^2; m_{\tilde{q}_{\text{L}/\text{R}}}^2, m_{\tilde{q}_{\text{L}/\text{R}}}^2) - 2B_0(m_q^2; m_3^2, m_{\tilde{q}_{\text{L}/\text{R}}}^2)\\ + 2(m_q^2 + m_3^2 - m_{\tilde{q}_{\text{L}/\text{R}}}^2)C_0(m_O^2,m_q^2,m_q^2; m_{\tilde{q}_{\text{L}/\text{R}}}^2,m_{\tilde{q}_{\text{L}/\text{R}}}^2,m_3^2)\end{multlined}\\
    \text{and}\ \ \ \mathcal{I}_{\text{L}/\text{R}}^{(2)}(O \to \bar{q}q) &= \begin{multlined}[t][10cm]2B_0(m_q^2; m_3^2, m_{\tilde{q}_{\text{L}/\text{R}}}^2) - 2B_0(m_O^2; m_3^2,m_3^2)\\ + (2m_q^2+2m_3^2-2m_{\tilde{q}_{\text{L}/\text{R}}}^2 - m_O^2)C_0(m_O^2,m_q^2,m_q^2; m_3^2,m_3^2,m_{\tilde{q}_{\text{L}/\text{R}}}^2).\end{multlined}
\end{align}
Now the sum in \eqref{eB7} is over shared (s)quark flavors.

\subsection*{$\mathcal{F}(o \to \bar{q}q)$: pseudoscalar decay to quarks}

The amplitude for this decay can be written for a single squark flavor, in close analogy with $\mathcal{M}^{(24)}(O \to \bar{q}q)$, as
\begin{align*}
\mathcal{M}(o \to \bar{q}q) = -3\ii\, \frac{g_3^3}{(4\pi)^2}\, m_3 m_q\, \bt{t}^a \int_{\text{F}} \left[\frac{1}{\Delta_2(y,z)}-\frac{1}{\Delta_4(y,z)}\right] \bar{u}(p_1,\sigma_1)\gamma_5 v(p_2,\sigma_2),
\end{align*}
where again
\begin{align*}
\int_{\text{F}} = \int_0^1 \d z \int_0^{1-z} \d y
\end{align*}
denotes the integral over the same functions \eqref{eB5} as for the scalar except with $m_O \to m_o$. The form factor $\mathcal{F}(o \to \bar{q}q)$ can be written accordingly as
\begin{align}\label{eB9}
\mathcal{F}(o \to \bar{q}q) = \sum_{\tilde{q}} \int_0^1 \d z \int_0^{1-z}\d y \left[\frac{1}{\Delta_2(y,z)} - \frac{1}{\Delta_4(y,z)}\right].
\end{align}
This form factor is also given in terms of Passarino-Veltman functions by
\begin{align}\label{eB10}
    \mathcal{F}(o \to \bar{q}q) = 16\ii \pi^2 \sum_{\tilde{q}} \left[\mathcal{I}_{\text{L}}(o \to \bar{q}q) - \mathcal{I}_{\text{R}}(o \to \bar{q}q)\right]
\end{align}
with
\begin{align}\label{eB11}
    \mathcal{I}_{\text{L}/\text{R}}(o \to \bar{q}q) = C_0(m_O^2,m_q^2,m_q^2;m_3^2,m_3^2,m_{\tilde{q}_{\text{L}/\text{R}}}^2).
\end{align}

\subsection*{$\mathcal{F}(O \to \tilde{t}_1 \bar{t} \tilde{\chi}_1^0)$: scalar three-body decay}

The integrand of the phase-space integral \eqref{e14} for this decay is given by
\begin{multline}\label{eB12}
\mathcal{F}(O \to \tilde{t}_1\bar{t}\tilde{\chi}_1^0) = \frac{M^2 - m_t^2 - m_{\tilde{\chi}_1^0}^2}{(M^2 - m_{\tilde{t}_1}^2)^2 - m_{\tilde{t}_1}^2 \Gamma_{\tilde{t}_1}^2}\\ \times \lambda^{1/2}(1,M^2m_O^{-2},m_{\tilde{t}_1}^2 m_O^{-2})\, \lambda^{1/2}(1,m_t^2 M^{-2},m_{\tilde{\chi}_1^0}^2 M^{-2}),
\end{multline}
where $\Gamma_{\tilde{t}_1}$ is the decay width of the light stop, and where the triangular function $\lambda$ is defined as
\begin{align}\label{eB13}
\lambda(x,y,z) = x^2 + y^2 + z^2 - 2(xy+xz+yz).
\end{align}
Analogous form factors for decays involving heavy stops or a bottom antiquark and lightest chargino are given by replacing $m_{\tilde{t}_1} \to m_{\tilde{t}_2}$ and $\Gamma_{\tilde{t}_1} \to \Gamma_{\tilde{t}_2}$ and/or $m_t \to m_b$ and $m_{\tilde{\chi}_1^0} \to m_{\tilde{\chi}_1^-}$.

\acknowledgments
This research was supported in part by the United States Department of Energy under grant DE-SC0011726. We are grateful to L. Darm\'{e} for useful discussion.

\bibliographystyle{Packages/JHEP}
\bibliography{Bibliography/bibliography.bib}

\end{document}